\documentclass[twocolumn,showpacs,preprintnumbers,amsmath,amssymb]{revtex4}
\usepackage{amsmath}
\usepackage{graphicx}
\usepackage{subfigure}
\usepackage{dcolumn}
\usepackage{color}

\usepackage{hyperref}
\begin{document}
\title{Motion of spinning particles around a polymer black hole in loop quantum gravity}

\date{\today}
\author{Ke Chen \footnote{E-mail: k.k.chen@foxmail.com}, Shao-Wen Wei \footnote{Corresponding author. E-mail: weishw@lzu.edu.cn}}

\affiliation{$^{1}$Lanzhou Center for Theoretical Physics, Key Laboratory of Theoretical Physics of Gansu Province, and Key Laboratory of Quantum Theory and Applications of MoE, Lanzhou University, Lanzhou, Gansu 730000, China,\\
 $^{2}$Institute of Theoretical Physics $\&$ Research Center of Gravitation, Lanzhou University, Lanzhou 730000, People's Republic of China}

\begin{abstract}
In the curved spacetime background, the trajectory of a spinning test particle will deviate from the geodesic. Using the effective potential method, we study the motion of a spinning test particle on the equatorial plane of a polymer black hole in loop quantum gravity described by the Mathisson-Papapetrou-Dixon equations with minimal spin-gravity interaction. We find that for the bounded orbits in the radial direction, the particle's motion is timelike when its spin is small. The radial range of the orbit and its eccentricity decrease with the loop quantum gravity parameter. However, when the particle takes a large enough spin, we observe an interesting phenomenon that the timelike and spacelike motions alternately appear while are separated by a critical radius. Outside the critical radius, the motion is timelike, however inside it is spacelike, and on the radius $r_c$ it is null. This property shares similarities with the event horizon radius. However, unlike the event horizon radius, the value of $r_c$ is related to both the particle's motion and the black hole parameters. To explore more observable effects of the loop quantum gravity parameter on the motion of the spinning particle, we focus our attention on the circular orbits, particularly the innermost stable circular orbits, near the black hole. The result shows that for the same spin, there are two different innermost stable circular orbits, one with a larger radius and the other with a smaller radius. Both the radii decrease as the loop quantum gravity parameter increases. More significantly, with the increase of the spin of the particle, the small innermost stable circular orbit transition from timelike to spacelike, while the one with large radius does not. Instead, it terminates at a certain value of spin. Furthermore, we identify the existence of a minimal spin value, below which the innermost stable circular orbits maintain a timelike motion irrespective of the loop quantum gravity parameter. All the results present the significant influences of the loop quantum gravity parameter on the motion of the spinning particles.
\end{abstract}



\maketitle

\section{Introduction}\label{sec:level1}
General relativity (GR) serves as a robust theoretical framework in astrophysics and cosmology, describing how matter impacts the structure of space and time, ultimately generating gravity. However, examinations of Einstein's field equations within GR have revealed the existence of singularities \cite{Penrose:1964wq,Hawking:1970zqf}, indicating that GR ceases to be applicable once spacetime curvature reaches the Planck scale. As an important conclusion of GR, black holes contain singularities and can produce many important and interesting phenomena, such as gravitational waves \cite{LIGOScientific:2016aoc}, gravitational lenses \cite{Fomalont:2009zg}, and accretion disks \cite{Abramowicz:2011xu}.

In order to overcome the singularity problem, various quantum gravity theories have been proposed, among which the most famous ones are loop quantum gravity (LQG) \cite{Bojowald:2007ky,Bojowald:2001xe,Ashtekar:2011ni,Olmedo:2017lvt}, string theory \cite{Cornalba:2002fi,Berkooz:2002je}, and non-commutative geometry \cite{Maceda:2003xr,Gorji:2014pka}. As a non-perturbative theory of quantum gravity, LQG has been extensively studied. In recent years, the polymer black hole, a quantum extension of the Schwarzschild black hole in the context of LQG, has been proposed \cite{Ashtekar:2020ifw,Bodendorfer:2019nvy,Bodendorfer:2019cyv}, in which the loop quantum effect is controlled by an uncertain parameter $A_{\lambda}$. The singularity of this black hole is replaced by a space-like transition surface, which smoothly connects an asymptotically Schwarzschild black hole to a white hole with mass $M_B$ and $M_W$, respectively. Employing with this polymer black hole in LQG, many issues, including the construction of rotating black holes \cite{Brahma:2020eos} and the formation of periodic orbits of test particles around this black hole \cite{Tu:2023xab}, have been explored. %Other related study on LQG black holes can be found in Refs. \cite{Islam:2022wck,Afrin:2022ztr,Vagnozzi:2022moj,KumarWalia:2022ddq,Yan:2022fkr,Alesci:2011wn,Chen:2011zzi,Dasgupta:2012nk,Barrau:2014yka,Hossenfelder:2012tc,Sahu:2015dea,Cruz:2015bcj,Pugliese:2020ivz,Pawlowski:2014nfa,Vaid:2012pr,Barrau:2011md,Modesto:2009ve,Yang:2023gas,Fu:2023drp,Addazi:2021xuf,Garcia-Chung:2020zyq,Garcia-Chung:2022pdy,Garcia-Chung:2023oul}.

The motion of test particles around black holes can provide valuable information about the properties of these astrophysical objects. The parameters of the black hole affect the trajectory of the test particles, which can lead to some observational effects. For example, the motion of circular or periodic orbits can result distinct feature in the gravitational waves produced by corresponding extreme-mass-ratio inspiral systems. The spinless particle within a polymer black hole in LQG has been extensively investigated due to the property that they move along the geodesics \cite{Tu:2023xab}.

However, for a spinning test particle, due to the spin-curvature force, its trajectory is non-geodesic \cite{Bardeen:1972fi,Hanson:1974qy}. By considering the polar-dipole approximation, the trajectory of a spinning particle in curved spacetime can be described by the well-known Mathisson-Papapetrou-Dixon (MPD) equations \cite{Mathisson:1937zz,Papapetrou:1951pa,Corinaldesi:1951pb,Dixon:1964cjb,Hojman1975,Hojman:1976kn}. Due to the existence of the spin-curvature force, the four-momentum of the spinning particle is not parallel to its four-velocity \cite{Hojman1975,Armaza:2015eha,Hojman:2012me}, where the four-momentum remains timelike along the trajectory, but the four-velocity may become superluminal \cite{Hojman1975,Armaza:2015eha,Hojman:2012me}. Some studies have showed that this superluminal behavior can be avoided if one takes into account nonminimal spin-gravitational interactions \cite{Deriglazov:2015zta,Deriglazov:2015wde,Ramirez:2017pmp,Deriglazov:2017jub}. The properties of a spinning test particle have also been studied in various black hole backgrounds \cite{Han:2010tp,Harms:2016ctx,Lukes-Gerakopoulos:2017vkj,Mukherjee:2018bsn,Zhang:2018eau,Pugliese:2013zma,Zhang:2017nhl,stuchlik1999equilibrium,Stuchlik:2006in,Plyatsko:2018oie,Han:2016cdh,Mukherjee:2018kju,Faye:2006gx,Witzany:2018ahb,Jefremov:2015gza,Nucamendi:2019qsn,Zhang:2016btg,Conde:2019juj,Liu:2019wvp,Suzuki:1997by,Han:2008zzf,Zhang:2020qew,Yang:2021chw}. Motivated by these results, in this paper, we focus on exploring the properties of the trajectories of a spin test particle in the context of a polymer black hole in LQG, especially for the innermost stable circular orbit (ISCO), which plays an important role in accretion disk theory \cite{Abramowicz:2011xu}.

The organization of this article is as follows. In Sec. \ref{2}, we briefly review the polymer black hole solution in LQG. In Sec. \ref{3}, we obtain the four-momentum and four-velocity of the spinning test particle by using the MPD equations. In Sec. \ref{4}, we employ the effective potential approach to investigate the orbits of the spinning test particle in detail. Finally, Sec. \ref{5} provides a brief summary and conclusion of our results. Through the paper, we use a geometrized unit system with $G = c = 1$, and adopt the metric convention $(-, +, +, +)$.

\section{Polymer black hole in loop quantum gravity}\label{2}
In this section, let us briefly review the polymer black hole. By solving the effective equations of LQG, the metric of this black hole can be written as \cite{Bodendorfer:2019nvy,Bodendorfer:2019cyv,Brahma:2020eos,Tu:2023xab}
\begin{equation}
ds^2=-8A_{\lambda}M_{B}^{2}\mathcal{A} \left( r \right) dt^2+\frac{dr^2}{8A_{\lambda}M_{B}^{2}\mathcal{A} \left( r \right)}+\mathcal{B} \left( r \right) d\Omega ^2,\label{eq1}
\end{equation}
which is a quantum extension of the Schwarzschild metric in LQG. The metric functions are defined in terms of radial variable as
\begin{align}
\mathcal{A} \left( r \right)&=\frac{1}{\mathcal{B} \left( r \right)}\left( 1+\frac{r^2}{8A_{\lambda}M_{B}^{2}} \right) \left( 1-\frac{2M_B}{\sqrt{8A_{\lambda}M_{B}^{2}+r^2}} \right),\label{eq2}\\
\mathcal{B} \left( r \right)&=\frac{512A_{\lambda}^{3}M_{B}^{4}M_{W}^{2}+\left( r+\sqrt{8A_{\lambda}M_{B}^{2}+r^2} \right) ^6}{8\sqrt{8A_{\lambda}M_{B}^{2}+r^2}\left( \sqrt{8A_{\lambda}M_{B}^{2}+r^2}+r \right) ^3},\label{eq3}
\end{align}
where $M_{B}$ and $M_{W}$ are two Dirac observables in the model. The  dimensionless parameter $A_{\lambda}$ is defined as $A_{\lambda}\equiv(\lambda_{k}/(M_{B}M_{W}))^{2/3}/2$, where the quantum parameter $\lambda_{k}$ is related to holonomy modifications in LQG \cite{Bodendorfer:2019nvy,Bodendorfer:2019cyv}.

The most important feature of this black hole is that the singularity of this black hole is replaced by a space-like transition surface, which smoothly connects an asymptotically Schwarzschild black hole to a white hole with mass $M_B$ and $M_W$, respectively, when the areal radius $\mathcal{B}(r)$ reaches a minimum value, which means the radius coordinate $r=0$ \cite{Bodendorfer:2019nvy,Bodendorfer:2019cyv,Brahma:2020eos}. This is similar to the quantum bounce in loop quantum cosmology (LQC). In this paper, we only focus on the case of the symmetric bounce in which $M_B=M_W=M$. Then the metric functions $\mathcal{A}(r)$ and $\mathcal{B}(r)$ can be rewritten in the form of
\begin{align}
\mathcal{A} \left( r \right) &=\frac{1}{\mathcal{B} \left( r \right)}\left( 1+\frac{r^2}{8A_{\lambda}M^2} \right) \left( 1-\frac{2M}{\sqrt{8A_{\lambda}M^2+r^2}} \right),\label{eq4}
\\
\mathcal{B} \left( r \right) &=2A_{\lambda}M^2+r^2.
\label{eq5}
\end{align}
By introducing the metric function $\tilde{\mathcal{A}}\left( r \right) =8A_{\lambda}M^2\mathcal{A} \left( r \right)$ \cite{Bodendorfer:2019nvy,Bodendorfer:2019cyv,Brahma:2020eos,Tu:2023xab}, the line element will be
\begin{equation}
	ds^2=-\tilde{\mathcal{A}}\left( r \right) dt^2+\frac{dr^2}{\tilde{\mathcal{A}}\left( r \right)}+\mathcal{B} \left( r \right) \left( d\theta ^2+\sin ^2\theta d\phi ^2 \right).\label{eq6}
\end{equation}
The event horizon can be obtained by solving $\tilde{\mathcal{A}}\left( r \right) =0$,
\begin{equation}
r_h=2M\sqrt{1-2A_{\lambda}}.\label{eq7}
\end{equation}
Obviously, the horizon does not exist if $A_{\lambda}>1/2$. When $A_{\lambda}=0$, the above metric exactly reduces to the Schwarzschild spacetime.

\section{MPD equation solutions\label{3}}

It is well known that, for a spinning test particle, its motion does not follow geodesics. However, under the polar-dipolar approximation, it can be described by the MPD equations \cite{Mathisson:1937zz,Papapetrou:1951pa,Corinaldesi:1951pb,Dixon:1964cjb,Hojman1975,Hojman:1976kn}
\begin{align}
\frac{DP^{\mu}}{D\lambda}&=-\frac{1}{2}R_{\nu \alpha \beta}^{\mu}u^{\nu}S^{\alpha \beta},\label{eq8}
\\
\frac{DS^{\mu \nu}}{D\lambda}&=P^{\mu}u^{\nu}-P^{\nu}u^{\mu},\label{eq9}
\end{align}
where $R^{\mu}_{\nu \alpha \beta}$ is the Riemann tensor, $S^{\mu \nu}$, $P^{\mu}$, and $u^{\mu}=dx^{\mu}/d\lambda$ are the spin tensor, four-momentum, and four-velocity of the particle, respectively, and $\lambda$ is an arbitrary affine parameter.

Note that in Eqs. (\ref{eq8}) and (\ref{eq9}), there are 13 variables in total for $S^{\mu \nu}$, $P^{\mu}$ and $u^{\mu}$. The remaining undetermined degrees of freedom are related to the center of mass of a spinning particle, which is observer-denpendent in relativity \cite{Costa:2014nta}. Therefore, we must add a condition called the ``spin supplementary condition'' \cite{Mathisson:1937zz,Dixon:1964cjb,tulczyjew1959motion,Frenkel:1926zz,Ohashi:2003we,Kyrian:2007zz} to determine the entire system. In our study, we adopt the Tulczyjew spin-supplementary condition \cite{tulczyjew1959motion}
\begin{equation}
P_{\mu}S^{\mu \nu}=0.\label{eq10}
\end{equation}
By making use of Eqs. (\ref{eq8}), (\ref{eq9}), and (\ref{eq10}), one can determine the mass $m$ and spin $s$ of the particle
\begin{align}
m^2&=-P^{\mu}P_{\mu},\label{eq11}
\\
s^2&=\frac{1}{2}S^{\mu \nu}S_{\mu \nu}.\label{eq12}
\end{align}
Now let us consider the motion of a spinning test particle on the equatorial plane and consider the case where the spin vector is perpendicular to the equatorial plane, i.e., $P^{\theta}=0$ and $S^{\mu \nu}=0$. Further using the spin-supplementary condition, we obtain the nonzero components
\begin{align}
S^{tr}&=-\frac{sP_{\phi}}{m}\sqrt{-\frac{g_{\theta \theta}}{g}}=-S^{rt},\label{eq13}
\\
S^{t\phi}&=\frac{sP_r}{m}\sqrt{-\frac{g_{\theta \theta}}{g}}=-S^{\phi t},\label{eq14}
\\
S^{r\phi}&=-\frac{sP_t}{m}\sqrt{-\frac{g_{\theta \theta}}{g}}=-S^{\phi r},\label{eq15}
\end{align}
of the spin tensor, where $g$ is the determinant of the metric.

Furthermore, it was found that there are two Killing vectors, $\xi ^{\mu}=\left( \partial _t \right) ^{\mu}$ and $\eta ^{\mu}=\left( \partial _{\phi} \right) ^{\mu}$, in the context of this black hole, allowing us to construct two constants, namely the energy and total angular momentum of the particle \cite{Hojman:1976kn,Zhang:2016btg},
\begin{align}
e&=-\xi ^{\mu}P_{\mu}+\frac{1}{2}S^{\mu \nu}\xi _{\mu ;\nu}=-P_t+\frac{s\sqrt{g_{\theta \theta}}}{2\sqrt{-g}}\left( -g_{tt,r}\frac{P_{\phi}}{m} \right) ,\label{eq16}
\\
j&=\eta ^{\mu}P_{\mu}-\frac{1}{2}S^{\mu \nu}\eta _{\mu ;\nu}=P_{\phi}+\frac{s\sqrt{g_{\theta \theta}}}{2\sqrt{-g}}\left( -g_{\phi \phi ,r}\frac{P_t}{m} \right) ,\label{eq17}
\end{align}
where Eqs. (\ref{eq13})-(\ref{eq15}) are used in the second step. The non-zero components of the momentum can be calculated as
\begin{align}
P_t&=-\frac{2m\gamma \left( 2\gamma \bar{e}+M^2\bar{s}\bar{j}\partial _rg_{tt} \right)}{4\gamma ^2+M^2\bar{s}^2\partial _rg_{tt}\partial _rg_{\phi \phi}},\label{eq18}
\\
P_{\phi}&=\frac{2m\gamma M\left( 2\gamma \bar{j}-\bar{e}\bar{s}\partial _rg_{33} \right)}{4\gamma ^2+M^2\bar{s}^2\partial _rg_{tt}\partial _rg_{\phi \phi}},\label{eq19}
\\
\left( P^r \right) ^2&=-\frac{m^2+g^{tt}\left( P_t \right) ^2+g^{\phi \phi}\left( P_{\phi} \right) ^2}{g_{rr}},\label{eq20}
\end{align}
where $\gamma =\sqrt{-g_{tt}g_{rr}g_{\phi \phi}}$. The quantities $\bar{e}=e/m$, $\bar{j}=j/mM$, and $\bar{s}=s/mM$ are dimensionless energy, orbital angular momentum, and spin angular momentum, respectively.

From Eq. (\ref{eq9}), we have
\begin{align}
\frac{DS^{tr}}{D\lambda}&=P^tu^r-P^ru^t,\label{eq21}
\\
\frac{DS^{t\phi}}{D\lambda}&=P^tu^{\phi}-P^{\phi}u^t.\label{eq22}
\end{align}
Since the choice of affine parameter does not change the motion of the particle, we choose $\lambda=t$ for simplicity, which gives $u^t=1$, and thus we have $u^r=\dot{r}$ and $u^{\phi}=\dot{\phi}$. Then substituting Eqs. (\ref{eq14}) and (\ref{eq15}) into (\ref{eq21}) and (\ref{eq22}), we obtain
\begin{align}
P^tu^r-P^r&=\frac{M\bar{s}}{2\gamma}g_{\phi \nu}R_{\alpha \beta \eta}^{\nu}u^{\alpha}S^{\beta \eta},\label{eq23}
\\
P^tu^{\phi}-P^{^{\phi}}&=-\frac{M\bar{s}}{2\gamma}g_{r\nu}R_{\alpha \beta \eta}^{\nu}u^{\alpha}S^{\beta \eta}.\label{eq24}
\end{align}
Note that the corresponding equations (\ref{eq23}) and (\ref{eq24}) given in Refs. \cite{Zhang:2017nhl,Zhang:2020qew,Yang:2021chw} are corrected. As a result, the expression for radial and tangential velocities are given by
\begin{align}
\dot{r}&=\frac{P^r+\frac{M\bar{s}}{2\gamma}R_{\phi t\alpha \beta}S^{\alpha \beta}}{P^t-\frac{M\bar{s}}{2\gamma}R_{\phi r\alpha \beta}S^{\alpha \beta}},\label{eq25}
\\
\dot{\phi}&=\frac{P^{\phi}-\frac{M\bar{s}}{2\gamma}R_{rt\alpha \beta}S^{\alpha \beta}}{P^t+\frac{M\bar{s}}{2\gamma}R_{r\phi \alpha \beta}S^{\alpha \beta }}.\label{eq26}
\end{align}
Nevertheless, despite the difference in the form of velocities, the conclusion for circular orbits remains valid given in Refs. \cite{Zhang:2017nhl,Zhang:2020qew,Yang:2021chw}, for the reason that $u^{\phi}$ returns to Eq. (\ref{eq26}) when $u^r$ vanishes. On the other hand, due to the existence of spin-curvature forces, the velocity and momentum are not parallel. The momentum always follows the trajectory in a timelike manner, but the velocity may exceed the speed of light. In order to deal with this issue, a superluminal constraint must be imposed on the velocity \cite{Zhang:2017nhl}, i.e.,
\begin{align}
u^{\mu}u_{\mu}=g_{tt}+g_{rr}\left( u^r \right) ^2+g_{\phi \phi}\left( u^{\phi} \right) ^2<0.\label{eq27}
\end{align}
The appearance of the superluminal problem is due to the fact that the MPD equation is constructed under the polar-dipole approximation. If one considers the non-minimal spin-gravitational interactions, the superluminal problem will be avoided \cite{Deriglazov:2015zta,Deriglazov:2015wde,Ramirez:2017pmp,Deriglazov:2017jub}.

It is worth noting that if we restore the values of $G$ and $c$, the spin $\bar{s}=sc/(GMm)$. Once the actual physical values of the mass $M$ of the background black hole, the mass $m$ of the test particle, and its spin angular momentum $s$ are known, $\bar{s}$ can be calculated accordingly. As an example, let's assume a Kerr black hole ($m\sim10^{32}kg$) orbiting around a supermassive black hole ($M\sim10^{36}kg$). In this case, its spin angular momentum is easily obtained as $s=aGm^2/c\sim2\times10^{47}akg\cdot m^2\cdot s^{-1}$, which allows us to calculate $\bar{s}=am/M$. Given the values of $a$ as 1/2 and 1 (an extreme black hole), we can approximate obtain the values of $\bar{s}$ as $5\times10^{-5}$ and $10^{-4}$. Such small values of $\bar{s}$ implie that the spin-gravity force is negligible and the trajectory of the test particle may always be timelike in this extreme mass ratio systems.

\section{ Motion of spinning test particles\label{4}}

In this section, by using the effective method, we would like to study the circular orbit and the ISCO of the spinning particles within the polymer black hole background in LQG.

\subsection{Effective potential and equatorial Motion}
\begin{figure}
	\subfigure[$A_{\lambda}=0.1,\bar{l}=3.5$]{
		\begin{minipage}{0.49\linewidth}
			\includegraphics[width=1\textwidth]{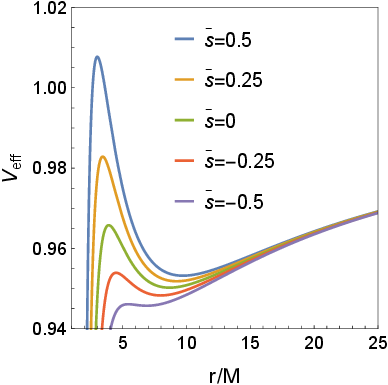}% Here is how to import EPS art
			\label{fig1a}
	\end{minipage}}\subfigure[$\bar{l}=3.5,\bar{s}=0.5$]{
		\begin{minipage}{0.49\linewidth}
			\includegraphics[width=1\textwidth]{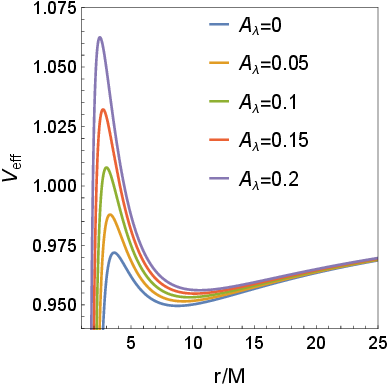}% Here is how to import EPS art
			\label{fig1b}
	\end{minipage}}
	\subfigure[$A_{\lambda}=0.1,\bar{s}=0.5$]{
		\begin{minipage}{0.49\linewidth}
			\includegraphics[width=1\textwidth]{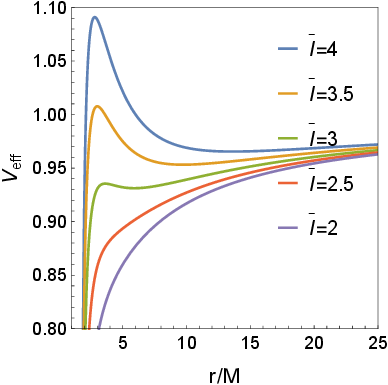}% Here is how to import EPS art
			\label{fig1c}
	\end{minipage}}\subfigure[$A_{\lambda}=0.1,\bar{s}=0.5$]{
		\begin{minipage}{0.49\linewidth}
			\includegraphics[width=1\textwidth]{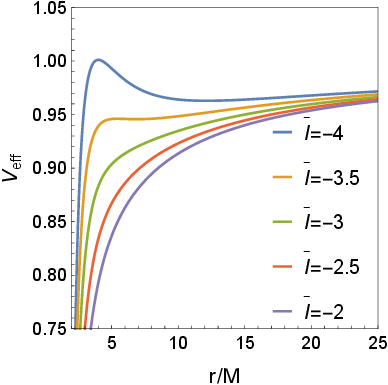}% Here is how to import EPS art
			\label{fig1d}
	\end{minipage}}
	\caption{\label{fig1} Effective potential $V_{eff}$ of a spinning particle for different parameters in the background of the polymer black holes.}
\end{figure}

The motion of particles in a central force field can be explored using an effective potential. From the expression of the radial velocity given in the previous section, it can be seen that $u^r$ is parallel to $P^r$. Thus, we can construct the effective potential with $P^r$ instead of $u^r$. The momentum $\left( P^r \right) ^2$ can be decomposed as
\begin{align}
\frac{\left( P^r \right) ^2}{m^2}&=\frac{1}{D}\left( A\bar{e}^2+B\bar{e}+C \right)\nonumber
\\
&=\frac{A}{D}\left( \bar{e}-\frac{-B+\sqrt{B^2-4AC}}{2A} \right) \nonumber
\\
&\times \left( \bar{e}-\frac{-B-\sqrt{B^2-4AC}}{2A} \right),\label{eq28}
\end{align}
where the functions $A$, $B$, $C$, and $D$ are
\begin{align}
A&=-16\gamma ^4g^{tt}-4\gamma ^2\bar{s}^2M^2g^{\phi \phi}\left( \partial _rg_{\phi \phi} \right) ^2,\label{eq29}
\\
B&=16\gamma ^3\bar{s}\bar{j}M^2\left( g^{\phi \phi}\partial _rg_{\phi \phi}-g^{tt}\partial _rg_{tt} \right), \label{eq30}
\\
C&=-16\gamma ^4\bar{j}^2M^2g^{\phi \phi}-4\gamma ^2\bar{s}^2\bar{j}^2M^4g^{tt}\left( \partial _rg_{tt} \right) ^2\nonumber
\\
&-\left( 4\gamma ^2+\bar{s}^2M^2\partial _rg_{tt}\partial _rg_{\phi \phi} \right) ^2,\label{eq31}
\\
D&=g_{rr}\left( 4\gamma ^2+\bar{s}^2M^2\partial _rg_{tt}\partial _rg_{\phi \phi} \right) ^2.\label{eq32}
\end{align}
Since the positive square root in Eq. (\ref{eq28}) represents the four-momentum pointing in the future direction, while the negative square root represents the four-momentum pointing in the past direction, we can choose the positive square root expression as the effective potential
\begin{align}
V_{eff}=\frac{-B+\sqrt{B^2-4AC}}{2A},
\end{align}
whose behavior is plotted in Fig. \ref{fig1} by varying different parameters.

As shown in Fig. \ref{fig1}, the effective potential has two extrema, with the maximum corresponding to an unstable circular orbit and the minimum corresponding to a stable one. In Fig. \ref{fig1a}, for the fixed LQG parameter $A_{\lambda}$ and orbital angular momentum $\bar{l}$, the effective potential increases with the spin angular momentum $\bar{s}$, which means that the trajectory of a spinning test particle is different from that of a non-spin particle in the same black hole background. In Fig. \ref{fig1b}, for the fixed $\bar{l}$ and $\bar{s}$, the effective potential increases with $A_{\lambda}$, and it will return back to that of a spinning particle in Schwarzschild spacetime, when $A_{\lambda}=0$. In Figs. \ref{fig1c} and \ref{fig1d}, for the fixed $A_{\lambda}$ and small $\bar{s}$, when $\bar{l}$ is aligned with $\bar{s}$, the effective potential increases with $\bar{l}$, while decreases with $\bar{l}$ if $\bar{l}$ is anti-aligned with $\bar{s}$. Moreover, it is evident that for the same spin, there are two different sets of stable and unstable circular orbits. To illustrate this point, we can calculate the radius of circular orbits $r$ and orbital angular momentum $\bar{l}$ for different spins.
\begin{figure}
	\begin{minipage}{0.99\linewidth}
		\includegraphics[width=0.5\textwidth]{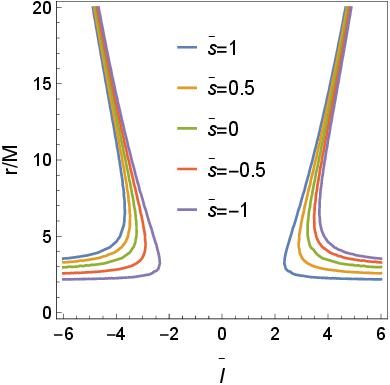}
		\caption{\label{fig2}The radius of the circular orbit as a function of the orbital angular momentum. We take $A_{\lambda}=0.1$.}
	\end{minipage}	
\end{figure}
\begin{figure*}[t]
	\subfigure[$A_{\lambda}=0$]{
		\begin{minipage}{0.32\linewidth}
			\includegraphics[width=0.9\textwidth]{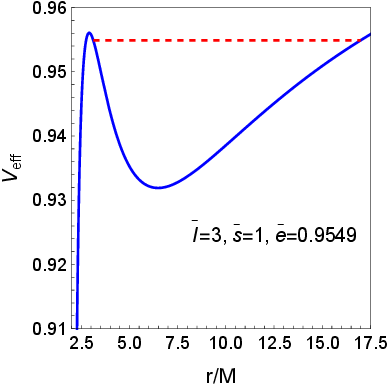}% Here is how to import EPS art
			\label{fig3a}
	\end{minipage}}\subfigure[$A_{\lambda}=0.2$]{
		\begin{minipage}{0.32\linewidth}
			\includegraphics[width=0.9\textwidth]{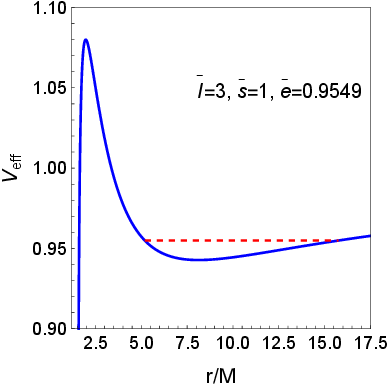}% Here is how to import EPS art
			\label{fig3b}
	\end{minipage}}\subfigure[$A_{\lambda}=0.5$]{
		\begin{minipage}{0.32\linewidth}
			\includegraphics[width=0.9\textwidth]{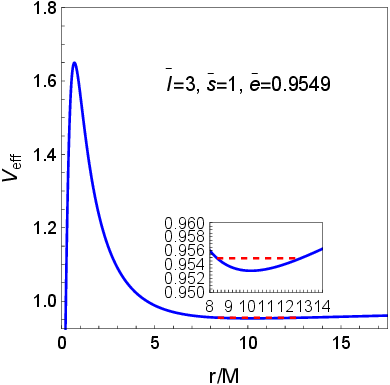}% Here is how to import EPS art
			\label{fig3c}
	\end{minipage}}
	\subfigure[$A_{\lambda}=0$]{
		\begin{minipage}{0.32\linewidth}
			\includegraphics[width=0.9\textwidth]{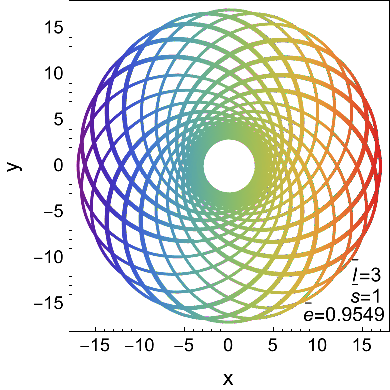}% Here is how to import EPS art
			\label{fig3d}
	\end{minipage}}\subfigure[$A_{\lambda}=0.2$]{
		\begin{minipage}{0.32\linewidth}
			\includegraphics[width=0.9\textwidth]{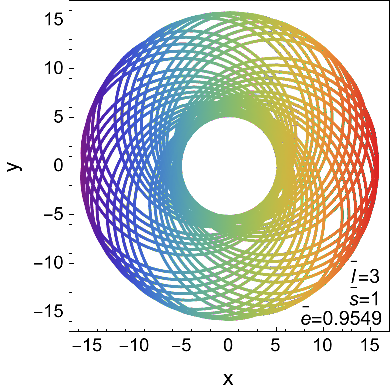}% Here is how to import EPS art
			\label{fig3e}
	\end{minipage}}\subfigure[$A_{\lambda}=0.5$]{
		\begin{minipage}{0.32\linewidth}
			\includegraphics[width=0.9\textwidth]{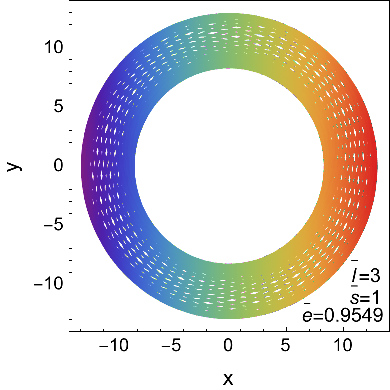}% Here is how to import EPS art
			\label{fig3f}
	\end{minipage}}
	\caption{\label{fig3} (a)-(c) The behaviors of the effective potential $V_{eff}$. The red dashed lines represent the energy $\bar{e}=0.9549$, and the particles move between the two radii $r_p$ and $r_a$ at the intersection of the dashed line and the effective potential. (d)-(e) The corresponding bounded trajectories.}
\end{figure*}

In Fig. \ref{fig2}, each spin value corresponds to two sets of stable and unstable circular orbits on the left and right sides, respectively. The inflection point of the right-hand curve represents the ISCO with a smaller radius (ISCOS), while that of the left-hand curve represents the innermost stable circular orbit with a larger radius (ISCOL). In fact, for the same spinning test particle, there is only one ISCO, but we found that when the spin is large, the circular orbit corresponding to the right inflection point (ISCO) may be superluminal, while the left one does not. Therefore, we define ISCOS and ISCOL separately to ensure that the ISCO is always physical. Detailed analysis will be presented in the next subsection. It is worth noting that the description of ``orbital angular momentum $\bar{l}$ and spin $\bar{s}$ being parallel or antiparallel" is not used here, instead the terms ``right-hand side" and ``left-hand side" are used. This is because when the spin is large enough, the values of $\bar{l}$ corresponding to the right-hand curve may become less than zero, thus it cannot be well described by the terms ``parallel" or ``antiparallel". In addition, multiple sets of stable and unstable circular orbits may appear for larger spin.
\begin{figure}
	\subfigure[]{
		\begin{minipage}{0.49\linewidth}
			\includegraphics[width=1\textwidth]{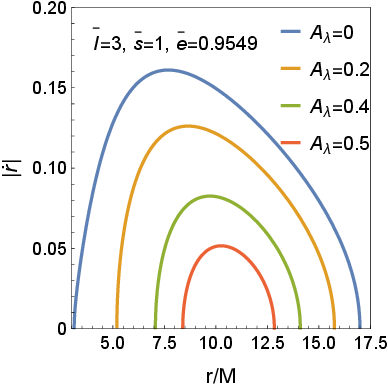}% Here is how to import EPS art
			\label{fig4a}
	\end{minipage}}\subfigure[]{
		\begin{minipage}{0.49\linewidth}
			\includegraphics[width=1\textwidth]{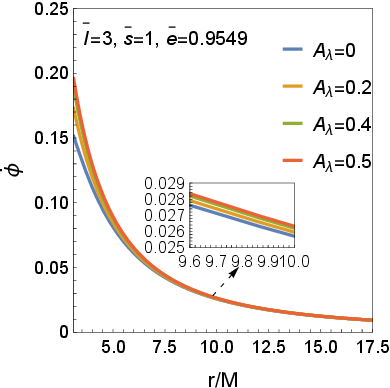}% Here is how to import EPS art
			\label{fig4b}
	\end{minipage}}
	\caption{\label{fig4}(a) The radial velocity $\dot{r}$ as a function of the radial coordinate $r$. (b) The angular velocity $\dot{\phi}$ as a function of the radial coordinate $r$.}
\end{figure}
\begin{figure}[h]
	\begin{minipage}{0.99\linewidth}
		\includegraphics[width=0.5\textwidth]{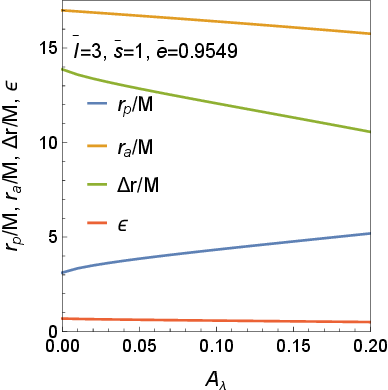}
		\caption{\label{fig5} The radii $r_{p}$ and $r_{a}$, their difference $r_{a}-r_{p}$, and the eccentricity of the bounded orbits as a function of the LOG $A_{\lambda}$.}
	\end{minipage}	
\end{figure}
\begin{figure*}[p]
	\subfigure[$\bar{e}=0.6731$]{
		\begin{minipage}{0.32\linewidth}
			\includegraphics[width=0.9\textwidth]{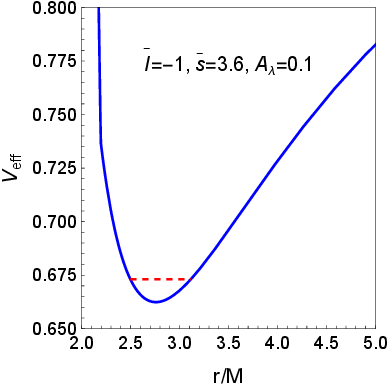}% Here is how to import EPS art
			\label{fig6a}
	\end{minipage}}\subfigure[$\bar{e}=0.6728$]{
		\begin{minipage}{0.32\linewidth}
			\includegraphics[width=0.9\textwidth]{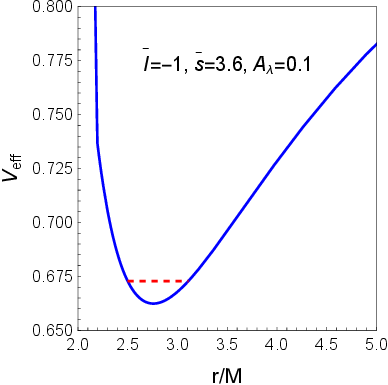}% Here is how to import EPS art
			\label{fig6b}
	\end{minipage}}\subfigure[$\bar{e}=0.6691$]{
		\begin{minipage}{0.32\linewidth}
			\includegraphics[width=0.9\textwidth]{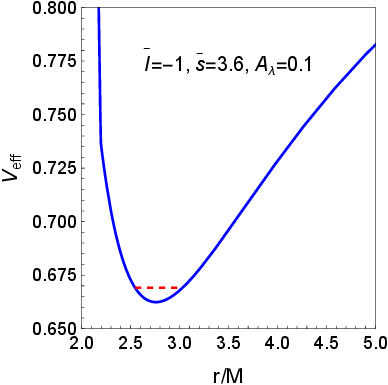}% Here is how to import EPS art
			\label{fig6c}
	\end{minipage}}
	\subfigure[$\bar{e}=0.6731$]{
		\begin{minipage}{0.32\linewidth}
			\includegraphics[width=0.9\textwidth]{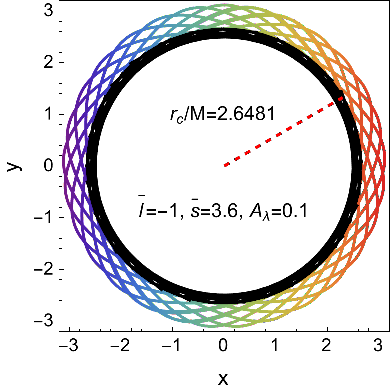}% Here is how to import EPS art
			\label{fig6d}
	\end{minipage}}\subfigure[$\bar{e}=0.6728$]{
		\begin{minipage}{0.32\linewidth}
			\includegraphics[width=0.9\textwidth]{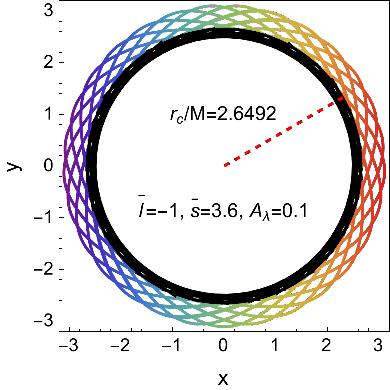}% Here is how to import EPS art
			\label{fig6e}
	\end{minipage}}\subfigure[$\bar{e}=0.6691$]{
		\begin{minipage}{0.32\linewidth}
			\includegraphics[width=0.9\textwidth]{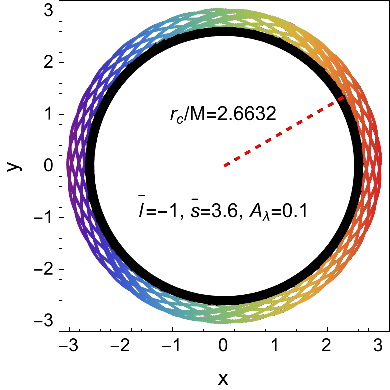}% Here is how to import EPS art
			\label{fig6f}
	\end{minipage}}
	\subfigure[$A_{\lambda}=0$]{
		\begin{minipage}{0.32\linewidth}
			\includegraphics[width=0.9\textwidth]{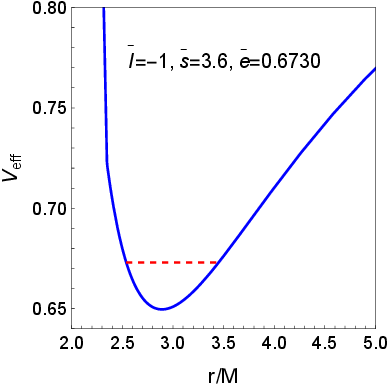}% Here is how to import EPS art
			\label{fig6g}
	\end{minipage}}\subfigure[$A_{\lambda}=0.1$]{
		\begin{minipage}{0.32\linewidth}
			\includegraphics[width=0.9\textwidth]{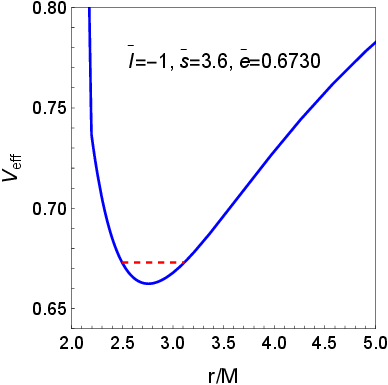}% Here is how to import EPS art
			\label{fig6h}
	\end{minipage}}\subfigure[$A_{\lambda}=0.15$]{
		\begin{minipage}{0.32\linewidth}
			\includegraphics[width=0.9\textwidth]{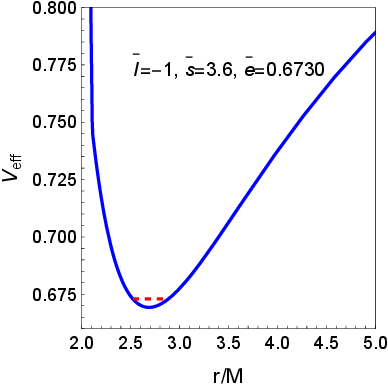}% Here is how to import EPS art
			\label{fig6i}
	\end{minipage}}
	\subfigure[$A_{\lambda}=0$]{
		\begin{minipage}{0.32\linewidth}
			\includegraphics[width=0.9\textwidth]{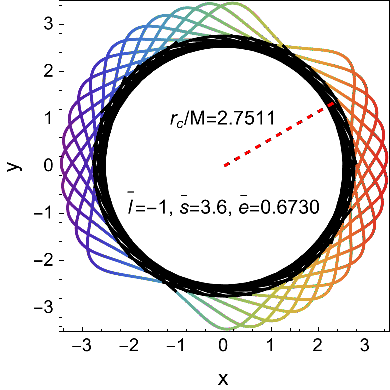}% Here is how to import EPS art
			\label{fig6j}
	\end{minipage}}\subfigure[$A_{\lambda}=0.1$]{
		\begin{minipage}{0.32\linewidth}
			\includegraphics[width=0.9\textwidth]{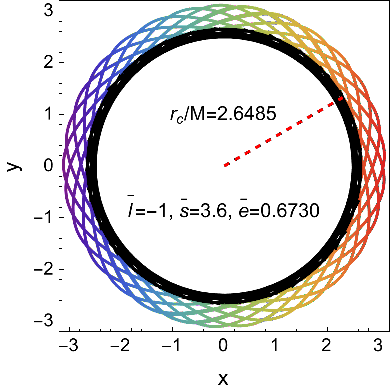}% Here is how to import EPS art
			\label{fig6k}
	\end{minipage}}\subfigure[$A_{\lambda}=0.15$]{
		\begin{minipage}{0.32\linewidth}
			\includegraphics[width=0.9\textwidth]{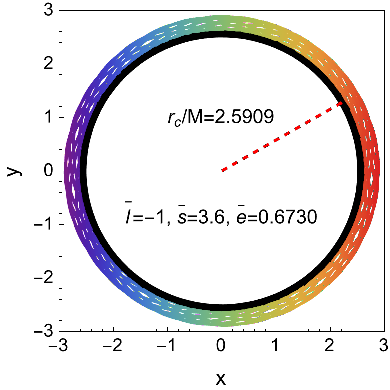}% Here is how to import EPS art
			\label{fig6l}
	\end{minipage}}
	\caption{\label{fig6}(a)-(c) and (g)-(i) The behaviors of the effective potentials. (d)-(f) and (j)-(l) The corresponding trajectories. The red lines in the effective potentials represent the fixed energy. $r_{c}$ denotes the critical radius of the timelike and spacelike trajectory motions.}
\end{figure*}

To better study the circular orbit motion of spin particles within this black hole background, let us first examine the case of spin particles moving in a non-circular orbit on the equatorial plane. As is well known, for a particle trapped in a potential well, it will be bounded between two radii $r_{p}$ (perihelion) and $r_{a}$ (aphelion) with $r_{a}>r_{b}$, as depicted in Fig. \ref{fig3}, where we take the energy $\bar{e}=0.9549$. In Fig. \ref{fig4}, as $A_{\lambda}$ increases, the radial velocity of the spin test particle decreases while the angular velocity increases, and the $r_{p}$ of the particle orbit increases while the $r_{a}$ decreases, which causes the orbit of the particle to change with $A_{\lambda}$, i.e., the range of motion of the particle narrows as the LQG effect increases. Fig. \ref{fig5} depicts the trend of orbit parameters with $A_{\lambda}$, where the shape of the orbit can be characterized by the eccentricity $\epsilon=(r_a-r_p)/(r_a+r_p)$, which decreases slightly as the LQG effect increases.

The above characteristics are universal for both large and small spins. In the case of small spin, the above types of orbits are timelike, but in the case of large spin, if the superluminal constraint is added, the particle's orbit can appear to be spacelike, as shown in Fig. \ref{fig6}. In Figs. \ref{fig6d}-\ref{fig6f} and \ref{fig6j}-\ref{fig6l}, the rainbow solid curves represent the timelike trajectories, while the black curves represent the spacelike trajectories. There is a critical radius $r_{c}$ between them. When a particle is released outside $r_{c}$, it can follow a rainbow trajectory that is both timelike and physical. Once crossing $r_{c}$, it becomes a spacelike and non-physical trajectory. At exactly $r_{c}$, its velocity is null. If the trajectory is required to be physical, the particle can only move outside the critical radius $r_{c}$.  As the particle approaches $r_{c}$, its velocity gradually approaches the speed of light. It is worth noting that the critical radius mentioned here is different from that mentioned in Ref. \cite{Deriglazov:2018zyp}, which is meaningless and will not occur during the motion of particles. However, according to numerical calculations, the critical radius here objectively exists. Interestingly, when $A_{\lambda}=0$, i.e., in the Schwarzschild background, the particle's trajectory exhibits this behavior as well, leading us to believe that this behavior also exists in other black hole backgrounds.

For simplicity, we explain this phenomenon using the example of a Schwarzschild black hole. In this context, Eqs. (\ref{eq25}) and (\ref{eq26}) become
\begin{align}
	\dot{r}&=\frac{P^r}{P^t}\label{eq34}
	\\
	\dot{\phi}&=\frac{P^{\phi}\left( r^3+2M^3\bar{s}^2 \right)}{P^t\left( r^3-M^3\bar{s}^2 \right)},\label{eq35}
\end{align}
which can be inserted into Eq. (\ref{eq27}) to form
\begin{align}
\left( \frac{ds}{dt} \right) ^2&=g_{tt}+g_{rr}\left( \dot{r} \right) ^2+g_{\phi \phi}\left( \dot{\phi} \right) ^2\nonumber
\\
&=\frac{g_{tt}\left( P^t \right) ^2\left( r^3-M^3\bar{s}^2 \right) ^2+g_{rr}\left( P^r \right) ^2\left( r^3-M^3\bar{s}^2 \right) ^2}{\left( P^t \right) ^2\left( r^3-M^3\bar{s}^2 \right) ^2}\nonumber
\\
&+\frac{g_{\phi \phi}\left( P^{\phi} \right) ^2\left( r^3+2M^3\bar{s}^2 \right) ^2}{\left( P^t \right) ^2\left( r^3-M^3\bar{s}^2 \right) ^2}.\label{eq36}
\end{align}
It is obvious that solving Eq. (\ref{eq36}) does not always yield a negative result. When we choose a suitable $\bar{l}$ and $\bar{e}$, as well as a relatively large $\bar{s}$, we can change the particle's trajectory from timelike to spacelike, thus resulting in a critical radius $r_{c}$. If we choose the proper time $\tau$ as the parameter, the square of the $t$-component of the 4-velocity is $(dt/d\tau)^2$, which is precisely the negative reciprocal of Eq. (\ref{eq36}). This result must be positive, leading to the same critical radius, where the $(dt/d\tau)^2$ diverges. Clearly, the critical radius $r_{c}$ prevents the trajectory from entering its inner region.

\begin{figure}
	\subfigure[]{
		\begin{minipage}{0.49\linewidth}
			\includegraphics[width=1\textwidth]{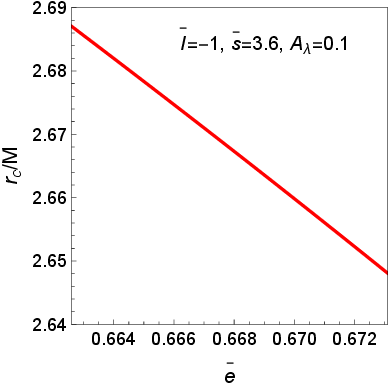}% Here is how to import EPS art
			\label{fig7a}
	\end{minipage}}\subfigure[]{
		\begin{minipage}{0.49\linewidth}
			\includegraphics[width=1\textwidth]{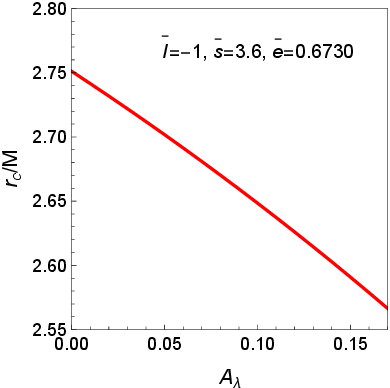}% Here is how to import EPS art
			\label{fig7b}
	\end{minipage}}
	\subfigure[]{
		\begin{minipage}{0.49\linewidth}
			\includegraphics[width=1\textwidth]{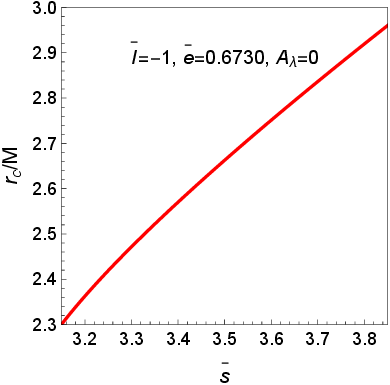}% Here is how to import EPS art
			\label{fig1c}
	\end{minipage}}\subfigure[]{
		\begin{minipage}{0.49\linewidth}
			\includegraphics[width=1\textwidth]{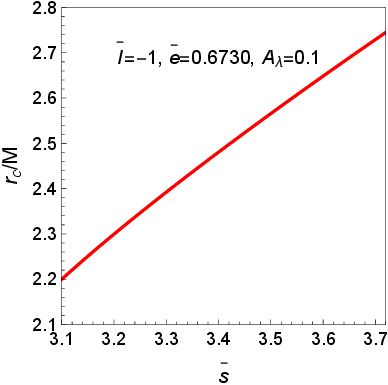}% Here is how to import EPS art
			\label{fig1d}
	\end{minipage}}
	\caption{\label{fig7} (a) The variation of critical radius $r_c$ with the energy $\bar{e}$. (b) The variation of critical radius $r_c$ with the LQG parameter $A_{\lambda}$. (c) and (d) The variation of critical radius $r_c$ with the spin $\bar{s}$.}
\end{figure}

From this perspective, although the critical radius $r_{c}$ acts like a horizon radius, which divides the region outside the radius from the region inside it, unlike the horizon radius, the size of $r_c$ is related to both the spin of the particle and the parameters of the black hole. In Fig. \ref{fig7}, it shows the variation of $r_{c}$ with $\bar{e}$, $A_{\lambda}$ and $\bar{s}$, indicating that $r_{c}$ decreases monotonically with $\bar{e}$ and $A_{\lambda}$, but increases with $\bar{s}$. It is worth noting that the appearance of the critical radius $r_{c}$ is attributed to the spin $s$, which makes the trajectory of the particle different from the geodesic. If a spinless particle moves along the geodesic, it will not exhibit the aforementioned behavior. Furthermore, the MPD equation that describes the motion of a spinning particle is not an exact equation. If one wants a more accurate description of the trajectory of a spinning particle, non-minimal spin-gravitational interactions should be taken into account \cite{Deriglazov:2015zta,Deriglazov:2015wde,Ramirez:2017pmp,Deriglazov:2017jub}.
\begin{table}
	\caption{\label{tab1}The values of LQG parameters $A_{\lambda}$ and corresponding to different total angular momentum $\bar{j}$ when the orbit is far away from the central black hole (893.285M, 24145.9M).}
	\begin{ruledtabular}
		\begin{tabular}{cc}	
			$\bar{j}$&$A_{\lambda}$\\
			\hline
			1&287.394\\
			2&286.893\\
			3&286.059\\
			38&46.9956\\
			39&34.1922\\
			40&21.0587\\
			41&7.59545\\
			41.53&0.326277\\
			\end{tabular}
	\end{ruledtabular}
\end{table}
\begin{table}
	\caption{\label{tab2}The values of LQG parameters $A_{\lambda}$ and corresponding to different total angular momentum $\bar{j}$ when the orbit is close to the central black hole (5M, 15M).}
	\begin{ruledtabular}
		\begin{tabular}{cc}
			$\bar{j}$&$A_{\lambda}$\\
			\hline
			1&1.4341\\
			2&0.960654\\
			3&0.36757\\
		\end{tabular}
	\end{ruledtabular}
\end{table}

In the above discussions, we have treated $A_{\lambda}$ as a variable. To obtain constraints on $A_{\lambda}$, one can use orbital data around the Sagittarius $\mathrm{A^{\star}}$. The star S2 is a good example that can be considered as a test particle around the central supermassive black hole. The perihelion and aphelion of the star S2 are both far away from the center, with distances of $893.285M$ and $24145.9M$ respectively \cite{Gillessen:2008qv}. Due to the nature of this extreme mass ratio, the magnitude of the spin is extremely small, being $10^{-6}$ \cite{Yang:2021chw}, which renders the spin-gravity force negligible. After calculation, two tables are presented, respectively showing the derived values of $A_{\lambda}$ with respect to the change in the total angular momentum $\bar{j}$ when the orbit is far from and close to the central black hole. From the two tables, it can be observed that if the particle moves away from the central black hole, the derived $A_{\lambda}$ will be very large (far greater than $1/2$), indicating that there will be no horizon radius. Additionally, when the value of $\bar{j}$ is relatively large, slight variations in it can lead to significant changes in $A_{\lambda}$, requiring precise control of $\bar{j}$ if we want to keep $A_{\lambda}$ within the range of less than $1/2$. However, when the particle moves near the central black hole, $A_{\lambda}$ will have a smaller value with less variability, making it easier to constrain the value of $A_{\lambda}$. These discussions imply that to accurately constrain the value of $A_{\lambda}$ in the EMRI system, orbits near the supermassive black hole should be taken into consideration. Of course, for the EMRI system, we can know the properties of the central supermassive black hole by detecting the gravitational waves it emits. Although the magnitude of the spin is very small due to the extreme mass ratio, small spin can produce considerable accumulated effect when we observe the secular evolution of EMRI, and can also make the phase of gravitational waves different from the spinless case \cite{Han:2010tp}. This property may be used for constraining the polymer black hole in LQG. However, different from spinless particles, the calculation of gravitational waves of spinning particles is more complicated and still under exploration \cite{Drummond:2023wqc}. The remaining discussions in this paper will not cover this aspect.
\begin{figure}
	\subfigure[$A_{\lambda}=0$]{
		\begin{minipage}{0.49\linewidth}
			\includegraphics[width=1\textwidth]{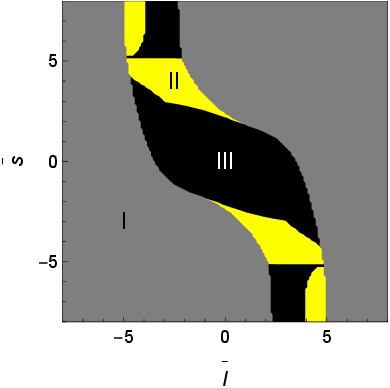}% Here is how to import EPS art
			\label{fig8a}
	\end{minipage}}\subfigure[$A_{\lambda}=0.1$]{
		\begin{minipage}{0.49\linewidth}
			\includegraphics[width=1\textwidth]{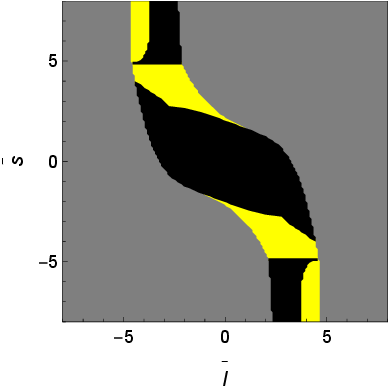}% Here is how to import EPS art
			\label{fig8b}
	\end{minipage}}
	\subfigure[$A_{\lambda}=0.2$]{
		\begin{minipage}{0.49\linewidth}
			\includegraphics[width=1\textwidth]{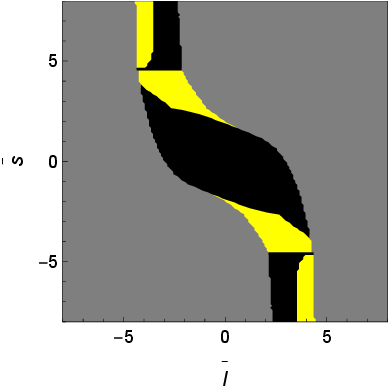}% Here is how to import EPS art
			\label{fig8c}
	\end{minipage}}\subfigure[$A_{\lambda}=0.3$]{
		\begin{minipage}{0.49\linewidth}
			\includegraphics[width=1\textwidth]{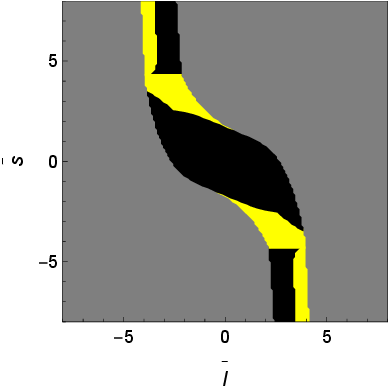}% Here is how to import EPS art
			\label{fig8d}
	\end{minipage}}
	\subfigure[$A_{\lambda}=0.4$]{
		\begin{minipage}{0.49\linewidth}
			\includegraphics[width=1\textwidth]{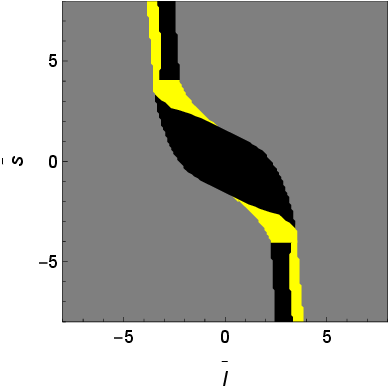}% Here is how to import EPS art
			\label{fig8e}
	\end{minipage}}\subfigure[$A_{\lambda}=0.5$]{
		\begin{minipage}{0.49\linewidth}
			\includegraphics[width=1\textwidth]{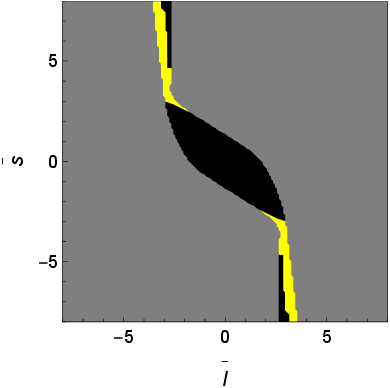}% Here is how to import EPS art
			\label{fig8f}
	\end{minipage}}
	\caption{\label{fig8}Plots of the region that the spinning test particle has a circular orbit in the ($\bar{s}-\bar{l}$) parameter space with $\bar{s}$ and $\bar{l}$ ranging from (-8, 8). The gray regions are for the particles that can have a stable timelike circular orbit, and the yellow regions are for the particles that have only the spacelike circular orbits. No circular orbit exists in the black color regions.}
\end{figure}

Next, let turn our attention to the circular orbits. By introducing the superluminal constraint and noting the energy $\bar{e}>0$, we numerically display the results in the ($\bar{s}-\bar{l}$) parameter space in Fig. \ref{fig8}. The particle can have timelike stable circular orbits in the gray region, only spacelike circular orbits in the yellow region, and no circular orbits in the black region. As $A_{\lambda}$ increases, the spacelike region and the region without circular orbits gradually shrink. It is evident that the timelike region is divided into two parts on the left and right. When $\bar{s}$ is greater than zero, the intersection of the gray region on the right with the black region at the center is an ISCOS, while the intersection of the gray region on the left with the black region at the center and the yellow region is an ISCOL. However, the situation is reversed when $\bar{s}$ is less than zero.

\begin{figure}
	\subfigure[$A_{\lambda}=0.1$]{
		\begin{minipage}{0.49\linewidth}
			\includegraphics[width=0.9\textwidth]{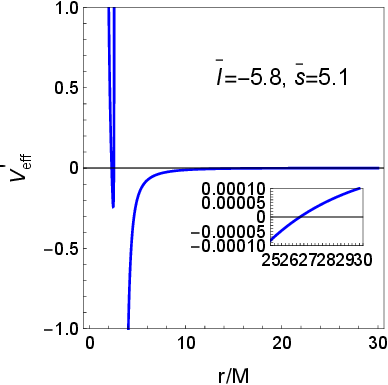}% Here is how to import EPS art
			\label{fig9a}
	\end{minipage}}\subfigure[$A_{\lambda}=0.1$]{
		\begin{minipage}{0.49\linewidth}
			\includegraphics[width=0.9\textwidth]{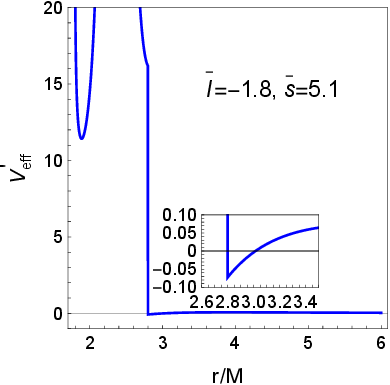}% Here is how to import EPS art
			\label{fig9b}
	\end{minipage}}
	\subfigure[$A_{\lambda}=0.2$]{
		\begin{minipage}{0.49\linewidth}
			\includegraphics[width=0.9\textwidth]{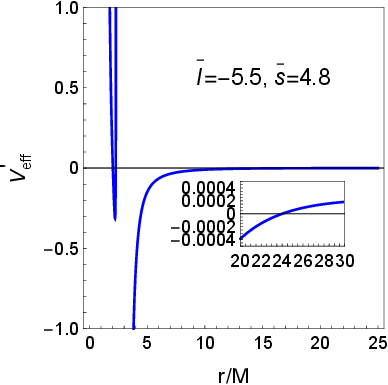}% Here is how to import EPS art
			\label{fig9c}
	\end{minipage}}\subfigure[$A_{\lambda}=0.2$]{
		\begin{minipage}{0.49\linewidth}
			\includegraphics[width=0.9\textwidth]{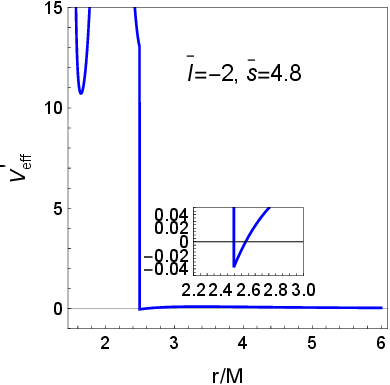}% Here is how to import EPS art
			\label{fig9d}
	\end{minipage}}
		\caption{\label{fig9}The first-order derivative of the effective potential $V_{eff}'$ as a function of the radial coordinate $r$ corresponding to different values of large spin $\bar{s}$ and $A_{\lambda}$.}
	\end{figure}

Considering the case of large spin, the effective potential will have multiple extrema and discontinuities in its first derivative. As shown in Figs. \ref{fig9a} and \ref{fig9c}, the first derivative of the effective potential has two zero points at smaller radii $r$, which means that the effective potential has both maximum and minimum values, indicating unstable and stable circular orbits. At larger radii $r$, there is only one zero point, indicating a single stable circular orbit. After a simple calculation, it is found that the inner two circular orbits are both spacelike, leaving only the outer single stable circular orbit. In Figs. \ref{fig9b} and \ref{fig9d}, the inner intersection of the first derivative of the effective potential with the zero axis is actually a discontinuity, similar to the situation where function $\left( x-1 \right)/\sqrt{\left( x-1 \right) ^2}$ has a singularity at $x=1$. Therefore, there is only one single stable circular orbit in this case as well.
	
\subsection{ISCO}

In this section, we aim at the ISCO of a spinning test particle. In the background of a polymer black hole in LQG, the presence of the ISCO is the combined result of the mass, spin of the particle, and the LQG effects.

For a particle moving around a black hole to form an ISCO, it must meet the following conditions:

\paragraph{the effective potential equals the energy}
\begin{equation}
V_{eff}=\bar{e},
\end{equation}
\paragraph{The first derivative of the effective potential is zero}
\begin{equation}
\frac{dV_{eff}}{dr}=0,
\end{equation}
\paragraph{The second derivative of the effective potential is zero}
\begin{equation}
\frac{d^2V_{eff}}{dr^2}=0.
\end{equation}
	\begin{figure}[b]
	\subfigure[$A_{\lambda}=0$]{
		\begin{minipage}{0.49\linewidth}
			\includegraphics[width=1\textwidth]{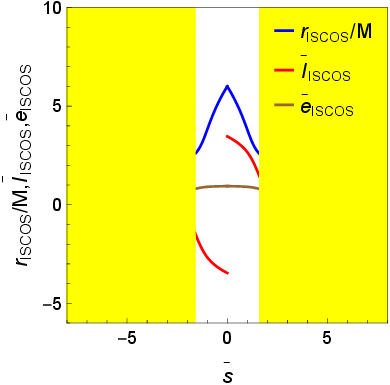}% Here is how to import EPS art
			\label{fig10a}
	\end{minipage}}\subfigure[$A_{\lambda}=0.2$]{
		\begin{minipage}{0.49\linewidth}
			\includegraphics[width=1\textwidth]{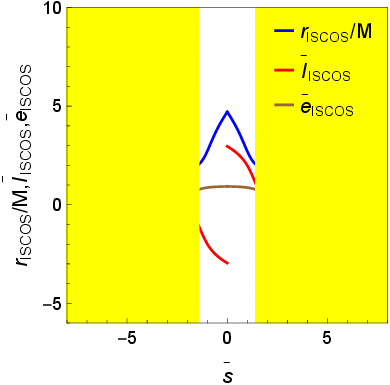}% Here is how to import EPS art
			\label{fig10b}
	\end{minipage}}
	\subfigure[$A_{\lambda}=0.3$]{
		\begin{minipage}{0.49\linewidth}
			\includegraphics[width=1\textwidth]{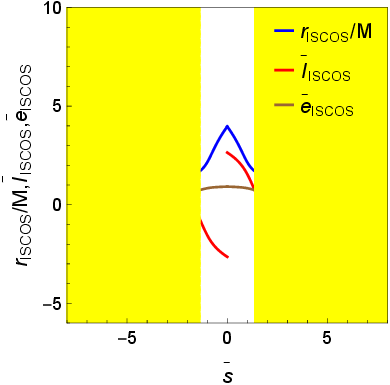}% Here is how to import EPS art
			\label{fig10c}
	\end{minipage}}\subfigure[$A_{\lambda}=0.5$]{
		\begin{minipage}{0.49\linewidth}
			\includegraphics[width=1\textwidth]{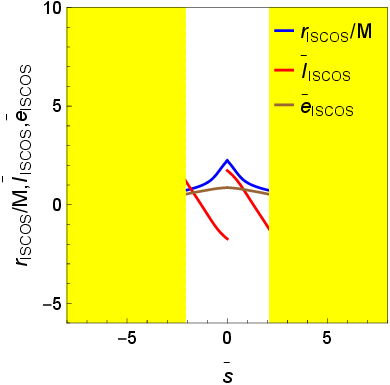}% Here is how to import EPS art
			\label{fig10d}
	\end{minipage}}
	\caption{\label{fig10}The variation of ISCOS with the change of the spin $\bar{s}$, where the yellow regions indicate that ISCOS motion is spacelike.}
\end{figure}

\begin{figure}
			\subfigure[$A_{\lambda}=0$]{
				\begin{minipage}{0.49\linewidth}
					\includegraphics[width=1\textwidth]{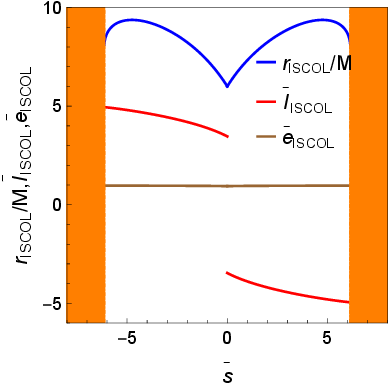}% Here is how to import EPS art
					\label{fig11a}
			\end{minipage}}\subfigure[$A_{\lambda}=0.2$]{
				\begin{minipage}{0.49\linewidth}
					\includegraphics[width=1\textwidth]{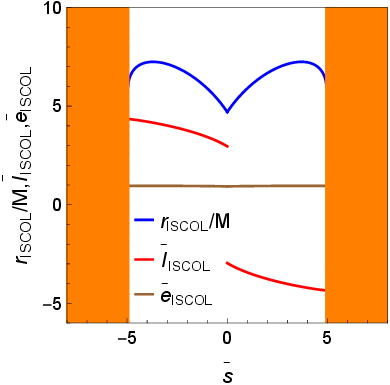}% Here is how to import EPS art
					\label{fig11b}
			\end{minipage}}
			\subfigure[$A_{\lambda}=0.3$]{
				\begin{minipage}{0.49\linewidth}
					\includegraphics[width=1\textwidth]{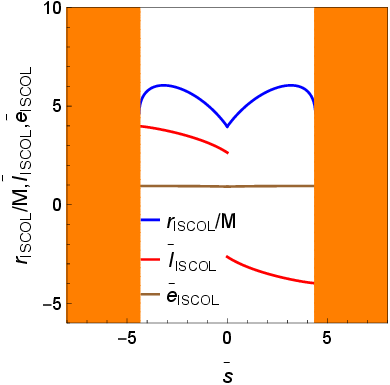}% Here is how to import EPS art
					\label{fig11c}
			\end{minipage}}\subfigure[$A_{\lambda}=0.5$]{
				\begin{minipage}{0.49\linewidth}
					\includegraphics[width=1\textwidth]{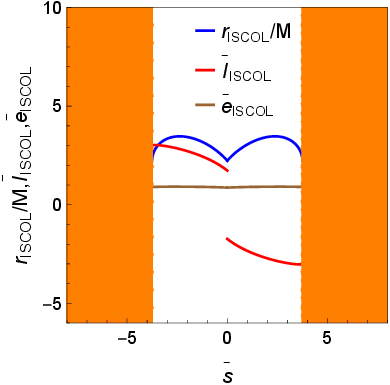}% Here is how to import EPS art
					\label{fig11d}
			\end{minipage}}
			\caption{\label{fig11}The variation of ISCOL with the change of the spin $\bar{s}$, where the orange regions indicate the absence of ISCOL.}
		\end{figure}

For a spin test particle, as depicted in Fig. \ref{fig2}, there exist an ISCOS and an ISCOL that both satisfy the above conditions within a certain range of spin. However, when taking ISCOS as ISCO, it is found that a spacelike situation occurs when the spin $\bar{s}$ is large enough, which has been observed in Refs. \cite{Zhang:2017nhl,Zhang:2020qew,Yang:2021chw}. When taking ISCOL, this situation does not occur. Figs. \ref{fig10} and \ref{fig11} show the variation of ISCOS and ISCOL with the spin $\bar{s}$.

In Figs. \ref{fig10} and \ref{fig11}, both ISCOS and ISCOL are symmetric with respect to the spin $\bar{s}$. In Fig. \ref{fig10}, the radius of ISCOS decreases with increasing $\bar{s}$, and it is spacelike in the yellow region. It is clear that the radius of ISCOS increases with the spin $\bar{s}$. When $A_{\lambda}=0$ and $\bar{s}=0$, the radius of ISCOS is exactly at $r=6M$, which is the radius of the ISCO for a spinless test particle in Schwarzschild spacetime. In Fig. \ref{fig11}, although ISCOL has no null region, it will terminate at a certain spin value $\bar{s}_e$, which decreases with $A_{\lambda}$. A simple calculation shows that when $A_{\lambda}=0$ (Schwarzschild spacetime), $\bar{s}_e\approx 6.1480$, and when $A_{\lambda}=0.5$ (no event horizon), $\bar{s}_e\approx 3.7666$. An interesting result is that as $A_{\lambda}$ increases, the critical spin $\bar{s}_c$ of the timelike and spacelike regions decreases first and then increases, as shown in Table \ref{tab3}. This feature suggests that there exists a minimum spin $\bar{s}_{min}$ such that the ISCO formed by particles with spin less than this $\bar{s}_{min}\approx1.3883$ is timelike for all $A_{\lambda}$.
\begin{table}
	\caption{\label{tab3}The values of spins $\bar{s}_c$ and $\bar{s}_e$ corresponding to different LQG parameters $A_{\lambda}$.}
	\begin{ruledtabular}
		\begin{tabular}{ccc}
			$A_{\lambda}$&$\bar{s}_c$&$\bar{s}_e$\\
			%\mbox{Three}&\mbox{Four}&\mbox{Five}\\
			\hline
			0&1.6518&6.1480\\
			0.1&1.5835& 5.5510\\
			0.2&1.4608&4.9586\\
			0.3&1.3960&4.4033\\
			0.4&1.4262&3.9627\\
			0.5&2.1419&3.7666\\
		\end{tabular}
	\end{ruledtabular}
\end{table}
\begin{figure}
	\subfigure[$\bar{s}=0$]{
		\begin{minipage}{0.49\linewidth}
			\includegraphics[width=1\textwidth]{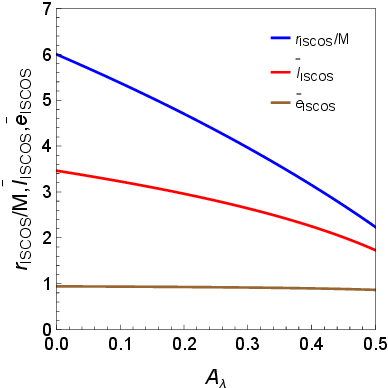}% Here is how to import EPS art
			\label{fig12a}
	\end{minipage}}\subfigure[$\bar{s}=1$]{
		\begin{minipage}{0.49\linewidth}
			\includegraphics[width=1\textwidth]{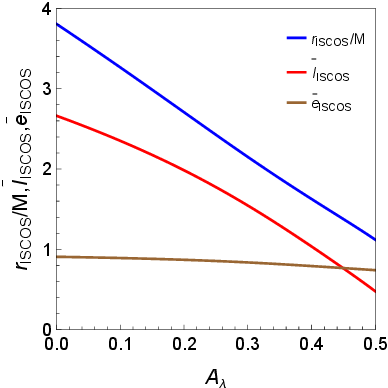}% Here is how to import EPS art
			\label{fig12b}
	\end{minipage}}
	\subfigure[$\bar{s}=1.4$]{
		\begin{minipage}{0.49\linewidth}
			\includegraphics[width=1\textwidth]{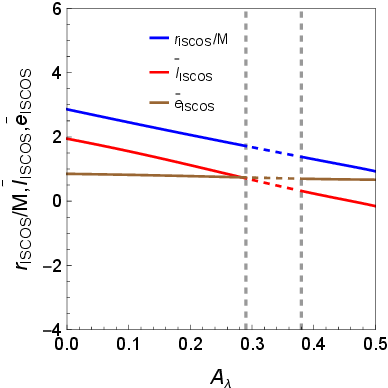}% Here is how to import EPS art
			\label{fig12c}
	\end{minipage}}\subfigure[$\bar{s}=3$]{
		\begin{minipage}{0.49\linewidth}
			\includegraphics[width=1\textwidth]{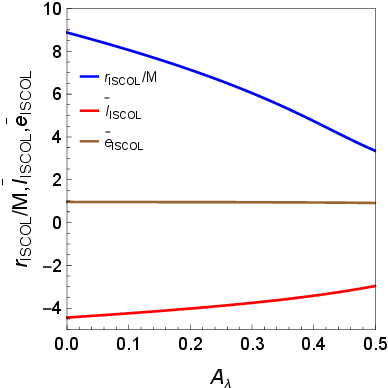}% Here is how to import EPS art
			\label{fig12d}
	\end{minipage}}
	%\subfigure[a=0.4]{
		%\begin{minipage}{0.16\linewidth}
		%\includegraphics[width=1\linewidth]{sfig/Vl3s1,a=0.4}% Here is how to import EPS art
		%\end{minipage}}\subfigure[a=0.5]{
		%\begin{minipage}{0.16\linewidth}
		%\includegraphics[width=1\linewidth]{sfig/Vl3s1,a=0.5}% Here is how to import EPS art
		%\end{minipage}}
		\caption{\label{fig12}The variation of ISCOS and ISCOL with $A_{\lambda}$ for different spin values $\bar{s}$. The dashed lines in (c) indicates that ISCOS is spacelike, and (d) shows the variation of ISCOL.}
	\end{figure}

Finally, we end this section with the figures of ISCOS and ISCOL varying with $A_{\lambda}$. Fig. \ref{fig12a} shows the ISCOS without spin, which is consistent with the result given in Ref. \cite{Tu:2023xab}. As $s$ increases, a null region appears. It is obvious that the ISCO of a polymer black hole in LQG is smaller than that of a Schwarzschild black hole, which means that a polymer black hole in LQG has a smaller inner edge of accretion disk compared to the Schwarzschild black hole with the same mass. Furthermore, the stronger the LQG effect is, the smaller the radius of the inner edge of accretion disk will be. Fig. \ref{fig12d} is the variation of ISCOL with $A_{\lambda}$, which does not exhibit a null region. It is clear that both the radii of ISCOS and ISCOL decrease as the LQG effect becomes stronger.

\section{ SUMMARY AND CONCLUSIONS\label{5}}

In this paper, we have investigated the trajectory of a spinning test particle on the equatorial plane of a polymer black hole in LQG by using the effective potential method. Due to the influence of spin-curvature force, the spinning test particle does not move along the geodesic. For simplicity, we only considered the parallel and anti-parallel cases of spin and orbital angular momentum. Moreover, we restricted our attention on the motion of spinning particle with a minimal spin-gravity interaction, which case can be well described by the MPD equations. Meanwhile, the superluminal constraint is also considered to ensure that the trajectory is physical.

Firstly, we obtained the effective potential of the test spinning particle through the property that the radial momentum and velocity are paralleled to each other. By confining the particle in the potential well, we obtained its trajectory on the equatorial plane. These trajectories are bounded between two radii $r_p$ and $r_a$. When adjusting the LQG parameter $A_{\lambda}$, we found that both the perihelion and angular velocity of the particle increase with $A_{\lambda}$, while the aphelion, eccentricity and radial decrease. These features are detailed examined when $A_{\lambda}$ varies.

Secondly, we observed a phenomenon that the timelike and spacelike trajectories of a particle alternate along its motion for the large spin. When the particle is released in the timelike region, it can undergo physical and timelike motion, and its speed approaches the speed of light continuously. The timelike and spacelike trajectories are separated by a critical radius $r_c$, which is similar to the nature of the horizon radius, but its size is related to both the spin of the particle and the parameters of the black hole. Interestingly, in the Schwarzschild background, the particle's trajectory exhibits this behavior as well, leading us to believe that this behavior is a universal phenomenon. It must be noted that this phenomenon is limited to the context of the MPD equation. If one wants a more accurate description of the trajectory of a spinning particle, non-minimal spin-gravitational interactions should be taken into account.

Then, in order to obtain the constraints on $A_{\lambda}$, we examined the orbits around the Sagittarius $\mathrm{A^{\star}}$. Due to the extreme mass ratio, the magnitude of spin angular momentum is very small, which makes the spin-gravity force negligible, providing convenience for our calculations. However, the large orbital radius makes it difficult for us to accurately control the range of $A_{\lambda}$, indicating that it is necessary to examine the orbits near the supermassive black hole if we want to accurately constrain $A_{\lambda}$ in the EMRI system. In addition, since the spin can change the phase of gravitational waves, detecting the gravitational waves from the EMRI system can also be used to constrain the polymer black hole in LQG. However, different from spinless particles, the calculation of gravitational waves of spinning particles is more complicated and still under exploration.

Finally, we investigated the ISCO of the spinning particle. To ensure that it remains physical as the spin $\bar{s}$ increases, we separated it into ISCOL with a larger radius and ISCOS with a smaller radius. The result shows that both the increase in the spin and the enhancement of LQG effects reduce the radius of ISCOS. As the spin increases, ISCOS exhibits a timelike region. This property implies that a polymer black hole in LQG has a smaller inner edge of accretion disk compared to the Schwarzschild black hole. On the contrary, ISCOL does not exhibit a timelike region, but terminates at a certain spin value, which will continue to decrease with the enhancement of the LQG effects.

In conclusion, we in this paper uncovered the LQG effects on the non-circular orbit, the circular orbits and ISCOs of the spinning particle within the polymer black hole. More detailed information of the LQG effects on the observed phenomena induced by the spinning particles are expected to be tested. With the improvement of the accuracy of future space gravitational wave detection and black hole shadow observation, we believe that our results will be useful to dig out the nature of the polymer black hole in LQG.

\section{ACKNOWLEDGMENTS}
This work was supported by the National Natural Science Foundation of China (Grants No. 12075103, No. 12247101). We would like to thank Dr. Yu-Peng Zhang for the useful discussions on the ($\bar{s}-\bar{l}$) parameter space.


\begin{thebibliography}{64}
%\cite{Penrose:1964wq}
\bibitem{Penrose:1964wq}
R.~Penrose,
{\em Gravitational collapse and space-time singularities},
Phys. Rev. Lett. \textbf{14}, 57 (1965).
%1717 citations counted in INSPIRE as of 27 Jan 2024

%\cite{Hawking:1970zqf}
\bibitem{Hawking:1970zqf}
S.~W.~Hawking and R.~Penrose,
{\em The Singularities of gravitational collapse and cosmology},
Proc. Roy. Soc. Lond. A \textbf{314}, 529 (1970).
%1214 citations counted in INSPIRE as of 27 Jan 2024

%\cite{LIGOScientific:2016aoc}
\bibitem{LIGOScientific:2016aoc}
B.~P.~Abbott \textit{et al.} [LIGO Scientific and Virgo],
{\em Observation of Gravitational Waves from a Binary Black Hole Merger},
Phys. Rev. Lett. \textbf{116}, 061102 (2016),
[\href{https://arxiv.org/abs/1602.03837}{arXiv:1602.03837} [gr-qc]].
%10634 citations counted in INSPIRE as of 27 Jan 2024

%\cite{Fomalont:2009zg}
\bibitem{Fomalont:2009zg}
E.~Fomalont, S.~Kopeikin, G.~Lanyi and J.~Benson,
{\em Progress in Measurements of the Gravitational Bending of Radio Waves Using the VLBA},
Astrophys. J. \textbf{699}, 1395 (2009),
[\href{https://arxiv.org/abs/0904.3992}{arXiv:0904.3992} [astro-ph.CO]].
%84 citations counted in INSPIRE as of 27 Jan 2024

%\cite{Abramowicz:2011xu}
\bibitem{Abramowicz:2011xu}
M.~A.~Abramowicz and P.~C.~Fragile,
{\em Foundations of Black Hole Accretion Disk Theory},
Living Rev. Rel. \textbf{16}, 1 (2013),
[\href{https://arxiv.org/abs/1104.5499}{arXiv:1104.5499} [astro-ph.HE]].
%359 citations counted in INSPIRE as of 27 Jan 2024

%\cite{Bojowald:2007ky}
\bibitem{Bojowald:2007ky}
M.~Bojowald,
{\em Singularities and Quantum Gravity},
AIP Conf. Proc. \textbf{910}, 294 (2007),
[\href{https://arxiv.org/abs/gr-qc/0702144}{arXiv:gr-qc/07021449} [gr-qc]].
%78 citations counted in INSPIRE as of 27 Jan 2024

%\cite{Bojowald:2001xe}
\bibitem{Bojowald:2001xe}
M.~Bojowald,
{\em Absence of singularity in loop quantum cosmology},
Phys. Rev. Lett. \textbf{86}, 5227 (2001),
[\href{https://arxiv.org/abs/gr-qc/0102069}{arXiv:gr-qc/0102069} [gr-qc]].
%613 citations counted in INSPIRE as of 27 Jan 2024

%\cite{Ashtekar:2011ni}
\bibitem{Ashtekar:2011ni}
A.~Ashtekar and P.~Singh,
{\em Loop Quantum Cosmology: A Status Report},
Class. Quant. Grav. \textbf{28}, 213001 (2011),
[\href{https://arxiv.org/abs/1108.0893}{arXiv:1108.0893} [gr-qc]].
%962 citations counted in INSPIRE as of 27 Jan 2024

%\cite{Olmedo:2017lvt}
\bibitem{Olmedo:2017lvt}
J.~Olmedo, S.~Saini and P.~Singh,
{\em From black holes to white holes: a quantum gravitational, symmetric bounce},
Class. Quant. Grav. \textbf{34}, 225011 (2017),
[\href{https://arxiv.org/abs/1707.07333}{arXiv:1707.07333} [gr-qc]].
%113 citations counted in INSPIRE as of 27 Jan 2024

%\cite{Cornalba:2002fi}
\bibitem{Cornalba:2002fi}
L.~Cornalba and M.~S.~Costa,
{\em A New cosmological scenario in string theory},
Phys. Rev. D \textbf{66}, 066001 (2002),
[\href{https://arxiv.org/abs/hep-th/0203031}{arXiv:hep-th/0203031} [hep-th]].
%294 citations counted in INSPIRE as of 27 Jan 2024

%\cite{Berkooz:2002je}
\bibitem{Berkooz:2002je}
M.~Berkooz, B.~Craps, D.~Kutasov and G.~Rajesh,
{\em Comments on cosmological singularities in string theory},
JHEP \textbf{03}, 031 (2003),
[\href{https://arxiv.org/abs/hep-th/0212215}{arXiv:hep-th/0212215} [hep-th]].
%111 citations counted in INSPIRE as of 27 Jan 2024

%\cite{Maceda:2003xr}
\bibitem{Maceda:2003xr}
M.~Maceda, J.~Madore, P.~Manousselis and G.~Zoupanos,
{\em Can noncommutativity resolve the big bang singularity?},
Eur. Phys. J. C \textbf{36}, 529 (2004),
[\href{https://arxiv.org/abs/hep-th/0306136}{arXiv:hep-th/0306136} [hep-th]].
%29 citations counted in INSPIRE as of 27 Jan 2024

%\cite{Gorji:2014pka}
\bibitem{Gorji:2014pka}
M.~A.~Gorji, K.~Nozari and B.~Vakili,
{\em Spacetime singularity resolution in Snyder noncommutative space},
Phys. Rev. D \textbf{89}, 084072 (2014)
[arXiv:1403.2623 [gr-qc]].
%22 citations counted in INSPIRE as of 27 Jan 2024

%\cite{Ashtekar:2020ifw}
\bibitem{Ashtekar:2020ifw}
A.~Ashtekar,
{\em Black Hole evaporation: A Perspective from Loop Quantum Gravity},
Universe \textbf{6}, 21 (2020),
[\href{https://arxiv.org/abs/2001.08833}{arXiv:2001.08833} [gr-qc]].
%49 citations counted in INSPIRE as of 27 Jan 2024

%\cite{Bodendorfer:2019nvy}
\bibitem{Bodendorfer:2019nvy}
N.~Bodendorfer, F.~M.~Mele and J.~M\"unch,
{\em (b,v)-type variables for black to white hole transitions in effective loop quantum gravity},
Phys. Lett. B \textbf{819}, 136390 (2021),
[\href{https://arxiv.org/abs/1911.12646}{arXiv:1911.12646} [gr-qc]].
%68 citations counted in INSPIRE as of 27 Jan 2024

%\cite{Bodendorfer:2019cyv}
\bibitem{Bodendorfer:2019cyv}
N.~Bodendorfer, F.~M.~Mele and J.~M\"unch,
{\em Effective Quantum Extended Spacetime of Polymer Schwarzschild Black Hole},
Class. Quant. Grav. \textbf{36}, 195015 (2019),
[\href{https://arxiv.org/abs/1902.04542}{arXiv:1902.04542} [gr-qc]].
%96 citations counted in INSPIRE as of 27 Jan 2024

%\cite{Brahma:2020eos}
\bibitem{Brahma:2020eos}
S.~Brahma, C.~Y.~Chen and D.~h.~Yeom,
{\em Testing Loop Quantum Gravity from Observational Consequences of Nonsingular Rotating Black Holes},
Phys. Rev. Lett. \textbf{126}, 181301 (2021),
[\href{https://arxiv.org/abs/2012.08785}{arXiv:2012.08785} [gr-qc]].
%66 citations counted in INSPIRE as of 27 Jan 2024

%\cite{Tu:2023xab}
\bibitem{Tu:2023xab}
Z.~Y.~Tu, T.~Zhu and A.~Wang,
{\em Periodic orbits and their gravitational wave radiations in a polymer black hole in loop quantum gravity},
Phys. Rev. D \textbf{108}, 2 (2023),
[\href{https://arxiv.org/abs/2304.14160}{arXiv:2304.14160} [gr-qc]].
%4 citations counted in INSPIRE as of 27 Jan 2024

%\cite{Bardeen:1972fi}
\bibitem{Bardeen:1972fi}
J.~M.~Bardeen, W.~H.~Press and S.~A.~Teukolsky,
{\em Rotating black holes: Locally nonrotating frames, energy extraction, and scalar synchrotron radiation},
Astrophys. J. \textbf{178}, 347 (1972).
%1802 citations counted in INSPIRE as of 27 Jan 2024

%\cite{Hanson:1974qy}
\bibitem{Hanson:1974qy}
A.~J.~Hanson and T.~Regge,
{\em The Relativistic Spherical Top},
Annals Phys. \textbf{87}, 498 (1974).
%303 citations counted in INSPIRE as of 27 Jan 2024

%\cite{Mathisson:1937zz}
\bibitem{Mathisson:1937zz}
M.~Mathisson,
{\em Neue mechanik materieller systemes},
Acta Phys. Polon. \textbf{6}, 163 (1937).
%508 citations counted in INSPIRE as of 27 Jan 2024

%\cite{Papapetrou:1951pa}
\bibitem{Papapetrou:1951pa}
A.~Papapetrou,
{\em Spinning test particles in general relativity. 1.},
Proc. Roy. Soc. Lond. A \textbf{209}, 248 (1951).
%835 citations counted in INSPIRE as of 27 Jan 2024

%\cite{Corinaldesi:1951pb}
\bibitem{Corinaldesi:1951pb}
E.~Corinaldesi and A.~Papapetrou,
{\em Spinning test particles in general relativity. 2.},
Proc. Roy. Soc. Lond. A \textbf{209}, 259 (1951).
%201 citations counted in INSPIRE as of 27 Jan 2024

%\cite{Dixon:1964cjb}
\bibitem{Dixon:1964cjb}
W.~G.~Dixon,
{\em A covariant multipole formalism for extended test bodies in general relativity},
Nuovo Cim. \textbf{34}, no.2, 317 (1964).
%257 citations counted in INSPIRE as of 27 Jan 2024

\bibitem{Hojman1975}
S.~A.~Hojman,
{\em Spinning Charged Test Particles in a Kerr-Newman Background},
Ph.D. thesis, Princeton University (1975).

%\cite{Hojman:1976kn}
\bibitem{Hojman:1976kn}
R.~Hojman and S.~Hojman,
{\em Spinning Charged Test Particles in a Kerr-Newman Background},
Phys. Rev. D \textbf{15}, 2724 (1977).
%70 citations counted in INSPIRE as of 27 Jan 2024

%\cite{Armaza:2015eha}
\bibitem{Armaza:2015eha}
C.~Armaza, M.~Ba\~nados and B.~Koch,
{\em Collisions of spinning massive particles in a Schwarzschild background},
Class. Quant. Grav. \textbf{33}, 105014 (2016),
[\href{https://arxiv.org/abs/1510.01223}{arXiv:1510.01223} [gr-qc]].
%50 citations counted in INSPIRE as of 27 Jan 2024

%\cite{Hojman:2012me}
\bibitem{Hojman:2012me}
S.~A.~Hojman and F.~A.~Asenjo,
{\em Can gravitation accelerate neutrinos?},
Class. Quant. Grav. \textbf{30}, 025008 (2013),
[\href{https://arxiv.org/abs/1203.5008}{arXiv:1203.5008} [physics.gen-ph]].
%48 citations counted in INSPIRE as of 27 Jan 2024

%\cite{Deriglazov:2015zta}
\bibitem{Deriglazov:2015zta}
A.~A.~Deriglazov and W.~G.~Ram\'\i{}rez,
{\em Mathisson\textendash{}Papapetrou\textendash{}Tulczyjew\textendash{}Dixon equations in ultra-relativistic regime and gravimagnetic moment},
Int. J. Mod. Phys. D \textbf{26}, 1750047 (2016),
[\href{https://arxiv.org/abs/1509.05357}{arXiv:1509.05357} [gr-qc]].
%40 citations counted in INSPIRE as of 27 Jan 2024

%\cite{Deriglazov:2015wde}
\bibitem{Deriglazov:2015wde}
A.~A.~Deriglazov and W.~G.~Ram\'\i{}rez,
{\em Ultrarelativistic Spinning Particle and a Rotating Body in External Fields},
Adv. High Energy Phys. \textbf{2016}, 1376016 (2016),
[\href{https://arxiv.org/abs/1511.00645}{arXiv:1511.00645} [gr-qc]].
%37 citations counted in INSPIRE as of 27 Jan 2024

%\cite{Ramirez:2017pmp}
\bibitem{Ramirez:2017pmp}
W.~G.~Ram\'\i{}rez and A.~A.~Deriglazov,
{\em Relativistic effects due to gravimagnetic moment of a rotating body},
Phys. Rev. D \textbf{96}, 124013 (2017),
[\href{https://arxiv.org/abs/1709.06894}{arXiv:1709.06894} [gr-qc]].
%20 citations counted in INSPIRE as of 27 Jan 2024

%\cite{Deriglazov:2017jub}
\bibitem{Deriglazov:2017jub}
A.~A.~Deriglazov and W.~Guzm\'an Ram\'\i{}rez,
{\em Recent progress on the description of relativistic spin: vector model of spinning particle and rotating body with gravimagnetic moment in General Relativity},
Adv. Math. Phys. \textbf{2017}, 7397159 (2017),
[\href{https://arxiv.org/abs/1710.07135}{arXiv:1710.07135} [gr-qc]].
%50 citations counted in INSPIRE as of 27 Jan 2024

%\cite{Han:2010tp}
\bibitem{Han:2010tp}
W.~B.~Han,
{\em Gravitational Radiations from a Spinning Compact Object around a supermassive Kerr black hole in circular orbit},
Phys. Rev. D \textbf{82}, 084013 (2010),
[\href{https://arxiv.org/abs/1008.3324}{arXiv:1008.3324} [gr-qc]].
%44 citations counted in INSPIRE as of 27 Jan 2024

%\cite{Harms:2016ctx}
\bibitem{Harms:2016ctx}
E.~Harms, G.~Lukes-Gerakopoulos, S.~Bernuzzi and A.~Nagar,
{\em Spinning test body orbiting around a Schwarzschild black hole: Circular dynamics and gravitational-wave fluxes},
Phys. Rev. D \textbf{94}, 104010 (2016),
[\href{https://arxiv.org/abs/1609.00356}{arXiv:1609.00356} [gr-qc]].
%67 citations counted in INSPIRE as of 27 Jan 2024

%\cite{Lukes-Gerakopoulos:2017vkj}
\bibitem{Lukes-Gerakopoulos:2017vkj}
G.~Lukes-Gerakopoulos, E.~Harms, S.~Bernuzzi and A.~Nagar,
{\em Spinning test-body orbiting around a Kerr black hole: circular dynamics and gravitational-wave fluxes},
Phys. Rev. D \textbf{96}, 064051 (2017),
[\href{https://arxiv.org/abs/1707.07537}{arXiv:1707.07537} [gr-qc]].
%50 citations counted in INSPIRE as of 27 Jan 2024

%\cite{Mukherjee:2018bsn}
\bibitem{Mukherjee:2018bsn}
S.~Mukherjee and K.~Rajesh Nayak,
{\em Off-equatorial stable circular orbits for spinning particles},
Phys. Rev. D \textbf{98}, 084023 (2018),
[\href{https://arxiv.org/abs/1804.06070}{arXiv:1804.06070} [gr-qc]].
%8 citations counted in INSPIRE as of 27 Jan 2024

%\cite{Zhang:2018eau}
\bibitem{Zhang:2018eau}
M.~Zhang and W.~B.~Liu,
{\em Innermost stable circular orbits of charged spinning test particles},
Phys. Lett. B \textbf{789}, 393 (2019),
[\href{https://arxiv.org/abs/1812.10115}{arXiv:1812.10115} [gr-qc]].
%24 citations counted in INSPIRE as of 27 Jan 2024

%\cite{Pugliese:2013zma}
\bibitem{Pugliese:2013zma}
D.~Pugliese, H.~Quevedo and R.~Ruffini,
{\em Equatorial circular orbits of neutral test particles in the Kerr-Newman spacetime},
Phys. Rev. D \textbf{88}, 024042 (2013),
[\href{https://arxiv.org/abs/1303.6250}{arXiv:1303.6250} [gr-qc]].
%88 citations counted in INSPIRE as of 27 Jan 2024

%\cite{Zhang:2017nhl}
\bibitem{Zhang:2017nhl}
Y.~P.~Zhang, S.~W.~Wei, W.~D.~Guo, T.~T.~Sui and Y.~X.~Liu,
{\em Innermost stable circular orbit of spinning particle in charged spinning black hole background},
Phys. Rev. D \textbf{97}, 084056 (2018),
[\href{https://arxiv.org/abs/1711.09361}{arXiv:1711.09361} [gr-qc]].
%32 citations counted in INSPIRE as of 27 Jan 2024

\bibitem{stuchlik1999equilibrium}
Zdenek.~Stuchl{\i}k,
{\em Equilibrium of spinning test particles in the Schwarzschild--de Sitter spacetimes},
Acta Phys. Slovaca \textbf{49}, 319 (1999).

%\cite{Stuchlik:2006in}
\bibitem{Stuchlik:2006in}
Z.~Stuchlik and J.~Kovar,
{\em Equilibrium conditions of spinning test particles in Kerr-de Sitter spacetimes},
Class. Quant. Grav. \textbf{23}, 3935 (2006),
[\href{https://arxiv.org/abs/gr-qc/0611153}{arXiv:gr-qc/0611153} [gr-qc]].
%18 citations counted in INSPIRE as of 27 Jan 2024

%\cite{Plyatsko:2018oie}
\bibitem{Plyatsko:2018oie}
R.~Plyatsko, V.~Panat and M.~Fenyk,
{\em Nonequatorial circular orbits of spinning particles in the Schwarzschild\textendash{}de Sitter background},
Gen. Rel. Grav. \textbf{50}, 150 (2018),
[\href{https://arxiv.org/abs/1811.01391}{arXiv:1811.01391} [gr-qc]].
%7 citations counted in INSPIRE as of 27 Jan 2024

%\cite{Han:2016cdh}
\bibitem{Han:2016cdh}
W.~B.~Han and R.~Cheng,
{\em Dynamics of extended bodies with spin-induced quadrupole in Kerr spacetime: generic orbits},
Gen. Rel. Grav. \textbf{49}, 48 (2017),
[\href{https://arxiv.org/abs/1611.07602}{arXiv:1611.07602} [gr-qc]].
%10 citations counted in INSPIRE as of 27 Jan 2024

%\cite{Mukherjee:2018kju}
\bibitem{Mukherjee:2018kju}
S.~Mukherjee,
{\em Collisional Penrose process with spinning particles},
Phys. Lett. B \textbf{778}, 54 (2018).
%23 citations counted in INSPIRE as of 27 Jan 2024

%\cite{Faye:2006gx}
\bibitem{Faye:2006gx}
G.~Faye, L.~Blanchet and A.~Buonanno,
{\em Higher-order spin effects in the dynamics of compact binaries. I. Equations of motion},
Phys. Rev. D \textbf{74}, 104033 (2006),
[\href{https://arxiv.org/abs/gr-qc/0605139}{arXiv:gr-qc/0605139} [gr-qc]].
%248 citations counted in INSPIRE as of 27 Jan 2024

%\cite{Witzany:2018ahb}
\bibitem{Witzany:2018ahb}
V.~Witzany, J.~Steinhoff and G.~Lukes-Gerakopoulos,
{\em Hamiltonians and canonical coordinates for spinning particles in curved space-time},
Class. Quant. Grav. \textbf{36}, 075003 (2019),
[\href{https://arxiv.org/abs/1808.06582}{arXiv:1808.06582} [gr-qc]].
%34 citations counted in INSPIRE as of 27 Jan 2024

%\cite{Jefremov:2015gza}
\bibitem{Jefremov:2015gza}
P.~I.~Jefremov, O.~Y.~Tsupko and G.~S.~Bisnovatyi-Kogan,
{\em Innermost stable circular orbits of spinning test particles in Schwarzschild and Kerr space-times},
Phys. Rev. D \textbf{91}, 124030 (2015),
[\href{https://arxiv.org/abs/1503.07060}{arXiv:1503.07060} [gr-qc]].
%89 citations counted in INSPIRE as of 27 Jan 2024

%\cite{Nucamendi:2019qsn}
\bibitem{Nucamendi:2019qsn}
U.~Nucamendi, R.~Becerril and P.~Sheoran,
{\em Bounds on spinning particles in their innermost stable circular orbits around rotating braneworld black hole},''
Eur. Phys. J. C \textbf{80}, 35 (2020),
[\href{https://arxiv.org/abs/1910.00156}{arXiv:1910.00156} [gr-qc]].
%17 citations counted in INSPIRE as of 27 Jan 2024

%\cite{Zhang:2016btg}
\bibitem{Zhang:2016btg}
Y.~P.~Zhang, B.~M.~Gu, S.~W.~Wei, J.~Yang and Y.~X.~Liu,
{\em Charged spinning black holes as accelerators of spinning particles},
Phys. Rev. D \textbf{94}, 124017 (2016),
[\href{https://arxiv.org/abs/1608.08705}{arXiv:1608.08705} [gr-qc]].
%39 citations counted in INSPIRE as of 27 Jan 2024

%\cite{Conde:2019juj}
\bibitem{Conde:2019juj}
C.~Conde, C.~Galvis and E.~Larra\~naga,
{\em Properties of the Innermost Stable Circular Orbit of a spinning particle moving in a rotating Maxwell-dilaton black hole background},
Phys. Rev. D \textbf{99}, 104059 (2019),
[\href{https://arxiv.org/abs/1905.01323}{arXiv:1905.01323} [gr-qc]].
%17 citations counted in INSPIRE as of 27 Jan 2024

%\cite{Liu:2019wvp}
\bibitem{Liu:2019wvp}
Y.~Liu and X.~Zhang,
{\em Maximal efficiency of the collisional Penrose process with spinning particles in Kerr-Sen black hole},
Eur. Phys. J. C \textbf{80}, 31 (2020),
[\href{https://arxiv.org/abs/1910.01872}{arXiv:1910.01872} [gr-qc]].
%13 citations counted in INSPIRE as of 27 Jan 2024

%\cite{Suzuki:1997by}
\bibitem{Suzuki:1997by}
S.~Suzuki and K.~i.~Maeda,
{\em Innermost stable circular orbit of a spinning particle in Kerr space-time},
Phys. Rev. D \textbf{58}, 023005 (1998),
[\href{https://arxiv.org/abs/gr-qc/9712095}{arXiv:gr-qc/9712095} [gr-qc]].
%80 citations counted in INSPIRE as of 27 Jan 2024

%\cite{Han:2008zzf}
\bibitem{Han:2008zzf}
W.~Han,
{\em Chaos and dynamics of spinning particles in Kerr spacetime},
Gen. Rel. Grav. \textbf{40}, 1831(2008),
[\href{https://arxiv.org/abs/1006.2229}{arXiv:1006.2229} [gr-qc]].
%65 citations counted in INSPIRE as of 27 Jan 2024

%\cite{Zhang:2020qew}
\bibitem{Zhang:2020qew}
Y.~P.~Zhang, S.~W.~Wei and Y.~X.~Liu,
{\em Spinning Test Particle in Four-Dimensional Einstein\textendash{}Gauss\textendash{}Bonnet Black Holes},
Universe \textbf{6}, 103 (2020),
[\href{https://arxiv.org/abs/2003.10960}{arXiv:2003.10960} [gr-qc]].
%99 citations counted in INSPIRE as of 27 Jan 2024

%\cite{Yang:2021chw}
\bibitem{Yang:2021chw}
K.~Yang, B.~M.~Gu and Y.~P.~Zhang,
{\em Motion of spinning particles around electrically charged black hole in Eddington-inspired Born\textendash{}Infeld gravity},
Eur. Phys. J. C \textbf{82}, 293 (2022),
[\href{https://arxiv.org/abs/2111.00864}{arXiv:2111.00864} [gr-qc]].
%5 citations counted in INSPIRE as of 27 Jan 2024

%\cite{Costa:2014nta}
\bibitem{Costa:2014nta}
L.~F.~O.~Costa and J.~Nat\'ario,
{\em Center of mass, spin supplementary conditions, and the momentum of spinning particles},
Fund. Theor. Phys. \textbf{179}, 215 (2015),
[\href{https://arxiv.org/abs/1410.6443}{arXiv:1410.6443} [gr-qc]].
%76 citations counted in INSPIRE as of 27 Jan 2024

\bibitem{tulczyjew1959motion}
W.~Tulczyjew,
{\em Motion of multipole particles in general relativity theory},
Acta Phys. Pol \textbf{18}, 94 (1959).

%\cite{Frenkel:1926zz}
\bibitem{Frenkel:1926zz}
J.~Frenkel,
{\em Die Elektrodynamik des rotierenden Elektrons},
Z. Phys. \textbf{37}, 243 (1926)
%281 citations counted in INSPIRE as of 27 Jan 2024

%\cite{Ohashi:2003we}
\bibitem{Ohashi:2003we}
A.~Ohashi,
{\em Multipole particle in relativity},
Phys. Rev. D \textbf{68}, 044009 (2003),
[\href{https://arxiv.org/abs/gr-qc/0306062}{arXiv:gr-qc/0306062} [gr-qc]].
%29 citations counted in INSPIRE as of 27 Jan 2024

%\cite{Kyrian:2007zz}
\bibitem{Kyrian:2007zz}
K.~Kyrian and O.~Semerak,
{\em Spinning test particles in a Kerr field},
Mon. Not. Roy. Astron. Soc. \textbf{382}, 1922 (2007).
%116 citations counted in INSPIRE as of 27 Jan 2024

%\cite{Deriglazov:2018zyp}
\bibitem{Deriglazov:2018zyp}
A.~A.~Deriglazov, W.~Guzm\'an Ram\'\i{}rez and P.~Rojas,
{\em Comment on ''Acceleration of particles to high energy via gravitational repulsion in the Schwarzschild field'' by C. H. McGruder III},
Astropart. Phys. \textbf{107}, 35 (2019),
[\href{https://arxiv.org/abs/1812.06832}{arXiv:1812.06832} [gr-qc]].
%3 citations counted in INSPIRE as of 02 Apr 2024

%\cite{Gillessen:2008qv}
\bibitem{Gillessen:2008qv}
S.~Gillessen, F.~Eisenhauer, S.~Trippe, T.~Alexander, R.~Genzel, F.~Martins and T.~Ott,
{\em Monitoring stellar orbits around the Massive Black Hole in the Galactic Center},
Astrophys. J. \textbf{692}, 1075 (2009),
[\href{https://arxiv.org/abs/0810.4674}{arXiv:0810.4674} [astro-ph]].
%1052 citations counted in INSPIRE as of 24 May 2024

%\cite{Han:2010tp}
\bibitem{Han:2010tp}
W.~B.~Han,
{\em Gravitational Radiations from a Spinning Compact Object around a supermassive Kerr black hole in circular orbit},
Phys. Rev. D \textbf{82}, 084013 (2010),
[\href{https://arxiv.org/abs/1008.3324}{arXiv:1008.3324} [gr-qc]].
%45 citations counted in INSPIRE as of 28 May 2024

%\cite{Drummond:2023wqc}
\bibitem{Drummond:2023wqc}
L.~V.~Drummond, P.~Lynch, A.~G.~Hanselman, D.~R.~Becker and S.~A.~Hughes,
{\em Extreme mass-ratio inspiral and waveforms for a spinning body into a Kerr black hole via osculating geodesics and near-identity transformations},
Phys. Rev. D \textbf{109}, 064030 (2024),
[\href{https://arxiv.org/abs/2310.08438}{arXiv:2310.08438} [gr-qc]].
%8 citations counted in INSPIRE as of 28 May 2024
\end{thebibliography}
\end{document}